\documentclass[twocolumn]{aastex63}
\usepackage{natbib}
\usepackage{color}
\usepackage{mathtools}
\usepackage{booktabs}
\usepackage{hyperref} 
\usepackage{mathtools}

\DeclarePairedDelimiter\ket{\lvert}{\rangle}
\DeclarePairedDelimiterX\braket[2]{\langle}{\rangle}{#1\,\delimsize\vert\,\mathopen{}#2}

\def	\cm		{\,{\rm {cm}}}
\def	\K		{\,{\rm K}}
\def	\g		{\,{\rm {g}}}
\def	\mum	{\,{\mu \rm{m}}}

\def \bea {\begin{eqnarray}}
\def \ena {\end{eqnarray}}

\def    \bB     {\bf  B}

\def	\bB	{\boldsymbol{B}}

\def 	\bE	{\boldsymbol{E}}
\def	\bF	{\boldsymbol{F}}

\def	\bJ	{\boldsymbol{J}} 
\def	\bk	{\boldsymbol{k}} 
 
\def	\bS	{\boldsymbol{S}} 
\def	\bn	{\boldsymbol{n}}
\def	\bp	{\boldsymbol{p}}
\def	\br	{\boldsymbol{r}}
\def	\bv	{\boldsymbol{v}}

\def	\cm	{\,{\rm cm}}

\def	\max	{\,{\rm max}}

\def	\erg	{\,{\rm erg}}
\def	\eV	{\,{\rm eV}\,}
\def	\g	{\,{\rm g}}

\def	\H	{{\rm H}}

\def	\s	{\,{\rm s}}

\def	\IP		{{\rm IP}}

\def    \ev     	{{\rm eV}}

\def	\yhat		{\hat{\bf y}}
\def	\zhat		{\hat{\bf z}}

\def    \bB     	{\boldsymbol{B}} 
\def    \bL     	{\boldsymbol{L}}
\def    \bn     	{\boldsymbol{n}} 
\def    \bv     	{\boldsymbol{v}} 
\def    \br     	{\boldsymbol{r}} 
\def    \bk     	{\boldsymbol{k}} 
\def  	\bF      	{\boldsymbol{F}}

\def	\bS			{\boldsymbol{S}} 
\def    \bP     	{\boldsymbol{P}}

\def	\bsigma		{\boldsymbol{\sigma}} 
\def	\bmu		{\boldsymbol{\mu}} 
 
\def	\bOmega		{\boldsymbol{\Omega}}

\usepackage{comment}
\begin{document}
\shorttitle{Role of aligned dust on chiral asymmetry}
\shortauthors{Hoang}
\title{Magnetically aligned grains as a ubiquitous source of spin-polarized electrons and chiral symmetry breaking agents}

\author{Thiem Hoang}
\affiliation{Korea Astronomy and Space Science Institute, Daejeon 34055, Republic of Korea} 
\email{thiemhoang@kasi.re.kr}
\affiliation{Department of Astronomy and Space Science, University of Science and Technology, 217 Gajeong-ro, Yuseong-gu, Daejeon, 34113, Republic of Korea}

\begin{abstract}
The unique biosignature of life on Earth is the homochirality of organic compounds such as amino acids, proteins, and sugars. High-energy spin-polarized (spin-up or spin-down) electrons (SPEs) from the $\beta$ decay of radioactive nuclei were proposed as a source of symmetry breaking, leading to homochirality; however, their exact role is much debated. Here we propose magnetically aligned dust grains as a new source of SPEs due to photoemission of electrons having aligned spins by the Barnett effect. For the interstellar UV radiation field of strength $G_{\rm UV}$, we found that the SPE emission rate is $\Gamma_{\rm pe}^{\rm SPE}\sim 10^{-14}G_{\rm UV}$ electrons per second per H, the fraction of spin-polarized to total photoelectrons is $\sim 10\%$, and the SPE yield (photoelectron number per UV photon) can reach $\sim 1\%$, using the modern theory of grain alignment. SPEs emitted from aligned grains could play an important role in chiral-induced spin selectivity-driven reduction chemistry in the icy grain mantles, producing an enantiomer excess of chiral molecules formed on the grain mantle. Finally, we suggest magnetically aligned grains could directly impact the enantioselectivity through the chiral-induced spin-selective adsorption effect and exchange interaction. We estimated the disalignment of electron spins and depolarization by elastic scattering using the Mott theory and found that these effects are negligible for low-energy SPEs, so that the spins of SPEs remain well aligned during their jurney through dust grains and the gas. Our proposed mechanism might explain the chiral asymmetry of prebiotic molecules detected in numerous comets, asteroids, and meteorites.
\end{abstract}


\section{Introduction}
Biological molecules such as amino acids and sugars have a unique geometrical property, so-called chirality (in Greek it means "hand"), discovered in 1848 by Louis Pasteur. A chiral molecule has two non-superimposable mirror images, called enantiomers. Chemical reactions typically produce a racemic mixture of equal amounts of left-handed and right-handed enantiomers (e.g., Miller-Urey experiment, \citealt{MillerUrey.1959}). Surprisingly, amino acids and proteins have only left-handed enantiomers and sugars, DNA and RNA have only right-handed enantiomers (see \citealt{Sparks.2015} for a review). Such a homochirality is the unique biosignature of life on Earth (e.g., \citealt{Bonner.1991}) and the most important biosignature of extraterrestrial life. 

Amino acids were discovered in several carbonaceous meteorites, including the Murchison meteorite \citep{Kvenvolden.1970,Cronin.1997,Pizzarello.1998}, the Murray meteorite \citep{Pizzarello.1998}, the Tagish Lake meteorite \citep{Glavin.2012}. The Rosetta mission detected the glycine in the coma of 67P/Churyumov-Gerasimenko \citep{Altwegg.2016} and the Ryugu asteroid \citep{Parker.2023,Oba.2023}. Most of those amino acids exhibit the L-enantiomer, the same as amino acids in biological molecules on Earth. Due to this similarity, a popular theory for the origin of life on Earth is that it arises from interstellar dust. The pressing question now is why there is the dominance of one enantiomer over the other (i.e., symmetry breaking or chiral asymmetry) of those biological molecules in meteorite (see \citealt{Sparks.2015,Brandenburg.2020} for reviews). If the symmetry breaking could be realized in astrophysical environments, it strengthens the hypothesis on the origin of life on Earth from space and paves the way for life elsewhere in the universe.

In the laboratory, it is found that UV circularly polarized photons can destroy one enantiomer, leaving only one enantiomer, a process called asymmetric photolysis \citep{Flores.1977}. Although it is not easy to destroy one enantiomer, the asymmetric photolysis produces a small enantiomer excess. This small initial enantiomer excess can be amplified by biological processes. CP is observed in star-forming regions \citep{Bailey.1998,Kwon.2016,Kwon.2018} and cometary coma \citep{Rosenbush.2007,Kolokolova.2016}. Therefore, the differential absorption of circularly polarized light by chiral molecules is a plausible mechanism producing the initial enantiomer excess (i.e., chiral asymmetry) and enantio-enrichment in the ISM \citep{Bailey.2001}.

\cite{Frank.1953} first suggested spontaneous symmetry breaking as a mechanism for the origin of homochirality from an initial racemic system due to random fluctuations (see more discussion in \citealt{Bonner.1991}). Soon after the discovery of parity violation in weak interactions \citep{LeeYang.1956,Wu.1957}, many studies suggested that the chiral asymmetry might arise through the preferential destruction of one enantiomer in a racemic mixture by spin-polarized electrons (SPEs) produced in the $\beta$ decay of radioactive nuclei. However, the induced chirality asymmetry was found to be rather weak \citep{Hegstrom.1980} and the exact role of parity violation is still debated (see \citealt{Bonner.1991} for a review). Moreover, unpolarized high-energy SPEs could dilute any chirality of molecules (see \citealt{Rosenberg.2019,Brandenburg.2020} for a review).
Recently, \cite{Globus.2020} and \cite{Globus.2021} revisited the effect of high-energy SPEs from parity violation and suggested that magnetically polarized CRs (e.g., muons) could induce initial chirality asymmetry in prebiotic molecules.

The last two decades have witnessed significant advances in understanding the origin of homochirality driven by the discovery that the transmission of electrons through chiral molecules is spin-dependent, an effect named chiral-induced spin selectivity (CISS) \citep{Ray.1999,Naaman.2012}. Numerous experiments established the key role of spin-polarized electrons and ferromagnetic surfaces as chiral agents (see reviews by \citealt{Naaman.2018,Rosenberg.2019}). For example, \cite{Rosenberg.2008} first demonstrated experimentally that low-energy ($\lesssim 10$ eV) SPEs resulting from irradiation of a magnetic substrate can induce chiral-selective chemistry in an adsorbed adlayer due to the CISS effect. Specifically, the authors found that the reaction cross-sections of a simple chiral molecule-(R)-or (S)-2-butanol (CH$_{3}$CHOHC$_{2}$H$_{5}$) under the irradiation of SPEs are spin-dependent, resulting in the ee of $\sim 10\%$ (see also \citealt{Rosenberg.2010}). Later on, \cite{Rosenberg.2015} found that SPEs produced by X-ray irradiation on a nonmagnetic gold surface that are transmitted through a chiral overlayer\footnote{Here the chiral overlayer acts as a filter that favors photoelectrons with a preferred spin state, so transmitted photoelectrons become spin-polarized.} induces chiral-selective chemistry in an adsorbed adlayer, which is caused by the different quantum yields for the reaction of SPEs with two enantiomers (see \citealt{Rosenberg.2019} for a review). Using this CISS paradigm, \cite{Ozturk.2022} suggested that SPEs induced by UV irradiation of magnetic deposits in the basin of an evaporative lake might induce CISS-driven reduction chemistry (CDRC) for prebiotic molecules in the lake-magnetite basin interface, resulting in the chiral asymmetry. Experimental study in \cite{Ozturk.2023a} showed that magnetic deposits act as chiral agents facilitating the homochiral enrichment of prebiotic compounds. Here, we propose interstellar dust grains aligned with magnetic fields as an important source of SPEs and chiral agents for interstellar chiral asymmetry. 

Interstellar dust is a magnetic material because observations show that more than $95\%$ of iron (one of the most abundant elements in the universe) is locked in dust \citep{Jenkins.2009}. Moreover, observations of starlight polarization \citep{Hall.1949,Hiltner.1949} and thermal dust polarization \citep{Planck.2015} revealed that dust grains in the interstellar medium (ISM) are asymmetric and efficiently aligned with ambient magnetic fields (see \citealt{Andersson.2015,LAH.2015} for reviews). In particular, high-resolution polarimetric observations by single-dish telescopes \citep{Pattle.2023ASPC} and interferometers like Atacama Large Millimeter Array (ALMA) \citep{Sadavoy.2019,Hull:2019hw,Gouellec.2020} reveal that dust grains are also aligned in very dense molecular clouds where young stars and planets are forming. The modern theory of grain alignment establishes that dust grains with embedded iron inclusions are efficiently aligned with the ambient magnetic field due to radiative torques and enhanced magnetic relaxation \citep{HoangLaz.2008,HoangLaz.2016,Hoangetal.2022}. 

Rapid rotation of interstellar magnetic grains due to gas- and radiation-dust interaction causes the alignment of electron spins along the rotation axis due to the Barnett effect \citep{Richardson.1908,Barnett.1915}. The Barnett effect is opposite to the popular Einstein-der Haas effect \citep{EinsteindeHaas.1915} that first showed the intrinsic connection between electron spin and macroscopic rotation of the solid body. The photoelectric effect on aligned grains induced by interstellar UV radiation can eject spin-polarized electrons spins aligned with the magnetic alignment axis due to angular momentum conservation \citep{Kessler.2013}. 

Ice mantles of interstellar dust grains play a crucial role in astrochemistry. Complex organic molecules (COMs, e.g., C$H_{3}$OH and C$_{2}$H$_{5}$OH), the building blocks of biological molecules such as amino acids and sugars, are believed to form in the ice mantle of dust grains from simple molecules such as H$_{2}0$, CO, HCN, and NH$_{3}$ \citep{Herbst:2009go,Caselli:2012fq}. Experimental studies by \cite{Bernstein.2002} and \cite{Caro.2002} independently demonstrated that amino acids (including glycine, alanine, and serine) could be formed naturally from UV photolysis of the analogs of interstellar ice (consisting of H$_2$O, HCN, NH$_{3}$, CH$_{3}$OH). The recent experiment by \cite{Esmaili.2018} showed that glycine can be formed by irradiation of low-energy electrons on interstellar analog of CO$_{2}$-CH$_{4}$-NH$_{3}$ ice (see \cite{Arumainayagam.2019} for a review). Therefore, SPEs from aligned dust grains could be important for formation of chiral molecules in the ice mantle of dust grains, in analogy to UVCPL \citep{Kessler.1998}. Moreover, magnetically aligned dust grains are equivalent to the ferromagnetic substrate in the lab, which would be a key chiral agent for chiral asymmetry. However, the efficiency of SPE emission depends on the grain alignment degree with the ambient magnetic field and the grain size distribution. We will first quantify the SPE emission yield of aligned grains using our modern theory of grain alignment and discuss implications for chiral asymmetry in this paper.

The paper structure is described as follows. In Section \ref{sec:dust_SPE} we present a new mechanism of SPE emission from aligned grains and a model for calculations of SPE photoemission based on modern grain alignment physics. Our numerical results for the rate and yield of SPE photoemission induced by interstellar UV radiation are shown in Section \ref{sec:results}. In Section \ref{sec:role_chiral} we study the effects of SPEs and aligned grains on chiral symmetry breaking. Discussion on the other radiation sources for SPEs, disalignment of spins of SPEs, and implications of our results for interstellar chiral asymmetry are presented in Section \ref{sec:discuss}. A summary of our main findings is presented in Section \ref{sec:summary}.

\section{Magnetically aligned grains and photoemission of spin-polarized electrons}\label{sec:dust_SPE}
Alignment of dust grains with the ambient magnetic field (i.e., magnetic alignment) is essential for the alignment of electron spins in space. Here, we briefly describe the dominant process inducing the magnetic alignment of grains.
\subsection{Grain magnetic moment from Barnett effect}
Interstellar dust grains are magnetic material due to the inclusion of iron atoms, in various forms, e.g., in the matrix of silicate (e.g., MgFeSiO$_{4}$) or as iron or iron-oxide nanoparticles (see e.g., \citealt{JonesSpitzer.1967,Draine.1996,HoangLaz.2016}. In this paper, we assume that silicate grains have oblate spheroidal shape and contain embedded iron clusters, which make them superparamagnetic material, as found by X-ray tomography study of dust grains by \cite{Hu.2021yit}.\footnote{This assumption is valid in a wide range of environments as constrained by polarimetric observations and grain alignment physics, from the diffuse ISM \citep{Draine.2022} to star- and planet-forming regions \citep{Giang.2023}.} Moreover, dust grains are rotating rapidly due to collisions with the gas \citep{Hoang.2010,HoangLaz.2016b} and radiative torques (RATs) caused by anisotropic radiation field \citep{Dolginov.1976,DraineWein.1997,LazHoang.2007,HoangLaz.2008}. 

A magnetic grain of zero-frequency susceptibility, $\chi(0)$, rotating with an angular velocity $\bOmega$ becomes magnetized via the Barnett effect \citep{Barnett.1915}, which is opposite to Einstein-de Haas effect. According to the Barnett effect, unpaired electrons within a rotating grain of angular velocity $\Omega$ are subject to an equivalent magnetic field given by
\bea
\bB_{\rm Bar}&=&-\frac{\bOmega}{|\gamma_{e}|}=-\frac{2m_{e}c}{eg_{e}}\bOmega,
\ena
where $\gamma_{e}=-g_{e}\mu_{B}/\hbar$ is the electron gyromagnetic ratio with $g_{e}\approx 2$ and $e$ the elementary charge, and $\mu_{B}=e\hbar/(2m_{e}c)$ is the magnetic moment of an electron (aka. Bohr magneton). 

\begin{figure}
\includegraphics[width=0.5\textwidth]{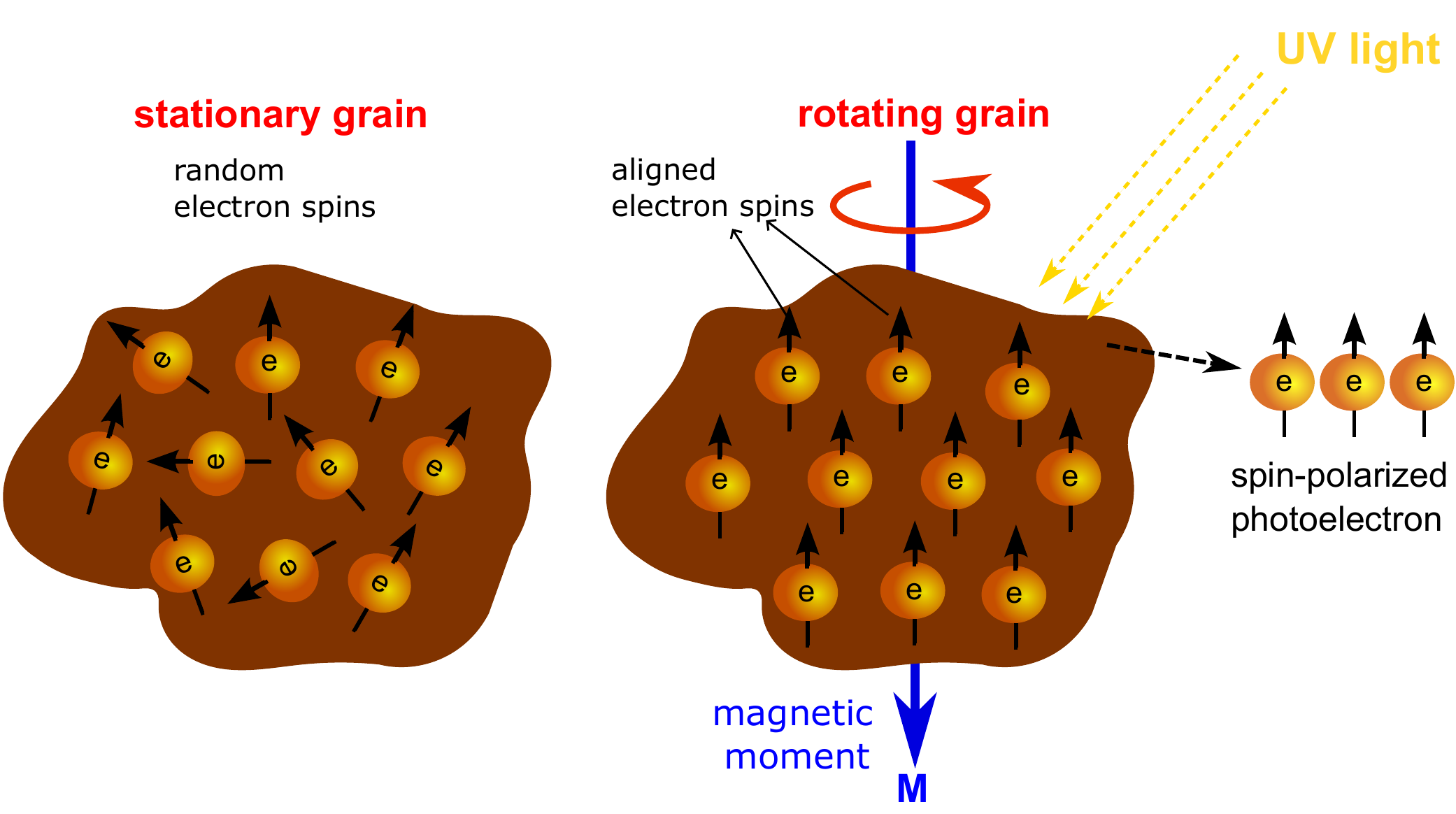}
\caption{Illustration of the magnetization of an irregularly-shaped grain by rotation (Barnett effect) and photoemission of spin-polarized electrons. Electron spins are random in the non-rotating grain (left) but are aligned along the rotation axis for the rotating grain (right). UV irradiation on the aligned rotating grain will produce spin-up or right-handed electrons.}
\label{fig:Barnett}
\end{figure}

The Barnett magnetic field acts to line up electron magnetic moments, which is equivalent to the alignment of electron magnetic moments within a body at rest by an external magnetic field. The resulting magnetic moment of the grain of volume $V=4\pi a^{3}/3$ is then given by
\bea
\bmu_{\rm Bar}&=& \chi(0)V{\bB}_{\rm Bar}=-\frac{\chi(0)V}{|\gamma_{e}|} \bOmega,\label{eq:muBar}
\ena
where the magnitude of $\chi(0)$ depends on magnetic properties of grains, such as paramagnetism, superparamagnetism, or ferro(ferri)magnetism (see e.g., \citealt{HoangLaz.2016,HoangLaz.2016b}). 

Figure \ref{fig:Barnett} illustrates the grain magnetization by rotation via the Barnett effect. Electron spins are random when the grain is at rest (left panel) and becomes aligned along the rotation axis (right panel). The photoelectric effect will eject electrons having random spin from the non-rotating grain (left) and aligned spin-up from the rotating grain.

\subsection{Grain alignment with the magnetic field}
Here, we briefly review the physical processes causing the alignment of magnetic grains with the magnetic field and present a model of their alignment efficiency.

The physical process of grain alignment with the ambient magnetic field is rather complicated, but it could be summarized as follows. First, fast internal relaxation by Barnett and inelastic relaxation \citep{Purcell.1979,LazEfroim.1999} and nuclear relaxation \citep{LazarianDraine.1999} induces the efficient alignment of the axis of maximum inertia with the grain angular momentum (i.e., internal alignment). Then, the Barnett magnetic moment allows the grain to interact with the ambient magnetic field via Larmor precession. Usually, the Larmor precession occurs much faster than the other timescales involved in grain alignment, such as gas randomization by gas collisions (see e.g., \citealt{Hoang.2021}), which makes the magnetic field the axis of grain alignment \citep{Lazarian.2007,LAH.2015,Hoangetal.2022}. Finally, radiative torques act to spin up grains to suprathermal rotation and align grains with the magnetic field \citep{DraineWein.1997,LazHoang.2007,HoangLaz.2008}. For grain alignment by RATs, grains can be aligned at an attractor with the grain angular momentum greater than its thermal angular momentum, named high-J attractor, and at an attractor point with the angular momentum comparable to the thermal value, named low-J attractor \citep{HoangLaz.2008}. The enhanced paramagnetic relaxation due to magnetic inclusions can further increase the magnetic alignment, which may make grains to be perfectly aligned by magnetically enhanced radiative torque (MRAT) mechanism \citep{HoangLaz.2016}.

Numerical simulations in \cite{HoangLaz.2008,HoangLaz.2016} show that if the RAT alignment has a high-J attractor point, then, large grains can be perfectly aligned because grains at low-J attractors would be randomized by gas collisions and eventually transported to more stable high-J attractors by RATs. On the other hand, grain shapes with low-J attractors would have negligible alignment due to gas randomization. For small grains, numerical simulations show that the alignment degree is rather small even in the presence of iron inclusions because grains rotate subthermally \citep{HoangLaz.2016}.

Therefore, the degree of grain alignment depends critically on the critical size above which grains can be aligned by RATs, denoted by $a_{\rm align}$. Accounting for the alignment of small grains, we can describe the size distribution of aligned grains, namely the grain alignment function as follows
\bea
f_{\rm align}(a)=f_{\rm min}+ (f_{\rm max}-f_{\rm min})\left[1-\exp\left(-\frac{a}{2a_{\rm align}}\right)^{3} \right],~~~\label{eq:falign}
\ena
where $f_{\rm min}$ describes the alignment degree of small grains of $a<a_{\rm align}$, including nanoparticles, and $f_{\rm max}$ is the maximum alignment degree of large grains of $a>a_{\rm align}$ \citep{HoangLaz.2016}. The alignment function $f_{\rm align}$ increases with $a$ and saturates at $f_{\rm max}$ for $a\gg a_{\rm align}$. The alignment function is narrower/broader when the alignment size $a_{\rm align}$ gets larger/smaller.

Numerical calculations in \cite{HoangLaz.2016b} showed that the degree of alignment for very small grains could reach $2-5\%$, and here we take $f_{\rm min}=0.025$. Although their alignment degree is much lower than larger grains, their dominance in the total surface area makes them a considerable source of SPEs. To account for the variation of grain efficiency by MRAT with dust magnetic susceptibility and local conditions (gas and radiation field), we vary the value of both $a_{\rm align}$ and $f_{\rm max}$ from $f_{\rm max}=1$ for perfect alignment to lower values for imperfect alignment.

According to the RAT alignment theory, grains with left and right helicity would be aligned with the high-J attractor point with $\bJ$ antiparallel and parallel to the magnetic field. Therefore, due to the Barnett effect, aligned grains of the left helicity will have electron spins anti-parallel to the magnetic field (called spin-down), while the right helicity grains will have electron spins parallel to the ambient magnetic field (spin-up). As a result, photoelectric emission from aligned grains will produce electrons with only one spin-up/spin-down state, namely spin-polarized electrons. 

\subsection{Photoelectric emission from aligned grains}
\subsubsection{Interstellar UV radiation}
Aligned grains are irradiated by diffuse interstellar radiation  (ISRF) from \cite{Mathis.1983}. The UV radiation spectrum of the ISRF can be approximately given by (e.g., \citealt{WeingartnerDraine.2001a}):
\bea
\nu u_{\nu}^{\rm MMP}= 
\left\{
\begin{array} {l l}
0 {\rm ~for~} h\nu>13.6 \ev\\
3.327\times 10^{-9} (h\nu/\ev)^{-4.4172}\erg\cm^{-3} \\
{\rm for~} 11.2\ev <h\nu<13.6\ev\\
8.463\times 10^{-13} (h\nu/\ev)^{-1}\erg\cm^{-3} \\
{\rm for~} 9.26\ev <h\nu<11.2\ev\\
2.055\times 10^{-14} (h\nu/\ev)^{0.6678}\erg\cm^{-3} \\
{\rm for~} 5.04\ev <h\nu<9.26\ev
\end{array}\right..
\label{eq:u_MMP}
\ena

To describe the variation of the local UV radiation field, we define $u_{\nu}^{\rm UV}=G_{\rm UV} u_{\nu}^{\rm MMP}$ where $G_{\rm UV}$ is the UV scaling factor. Equation (\ref{eq:Jpe}) implies that the photoelectron emission rate scales as $G_{\rm UV}$, and the fraction of SPE, $f_{\rm SPE}$ is independent of $G_{\rm UV}$. 

\subsubsection{Photoelectric effect and photoelectric yield}
Irradiation of dust grains aligned with the ambient magnetic field by unpolarized UV radiation would eject spin-polarized electrons with spins directed along the magnetic field. 

Let $Y(a,\nu)$ be the photoelectric yield of a grain of size $a$ induced by a photon of frequency $\nu$. The rate of photoelectron emission of primary electrons from one grain is
\bea
J_{\rm pe}(a)=\int_{\nu_{\rm pet}}^{\infty} Y(a,\nu)\pi a^{2}Q_{\rm abs} \frac{cu_{\nu}}{h\nu} d\nu,\label{eq:Jpe}
\ena
where $Q_{\rm abs}$ is the absorption efficiency, $\nu_{\rm pet}$ is the frequency threshold required for the photoelectric effect, which is determined by the ionization potential (IP), i.e., $h\nu_{\rm pet}=\IP$, and  $u_{\nu}$ is the specific energy density of the radiation field. Here we take the ionization potential $\IP=W=8$ eV for silicate grains (see \citealt{WeingartnerDraine.2001a}).

We calculate the photoelectric yield of grains irradiated by energetic photons for the different grain sizes using the method in \cite{Hoang.2015} (see \citealt{WeingartnerDraine.2001a}). For the interstellar UV radiation, we take the absorption efficiency $Q_{\rm abs}\approx 1$ (see Figure 16 in \citealt{Hoang.2015}).

\begin{figure}
\includegraphics[width=0.5\textwidth]{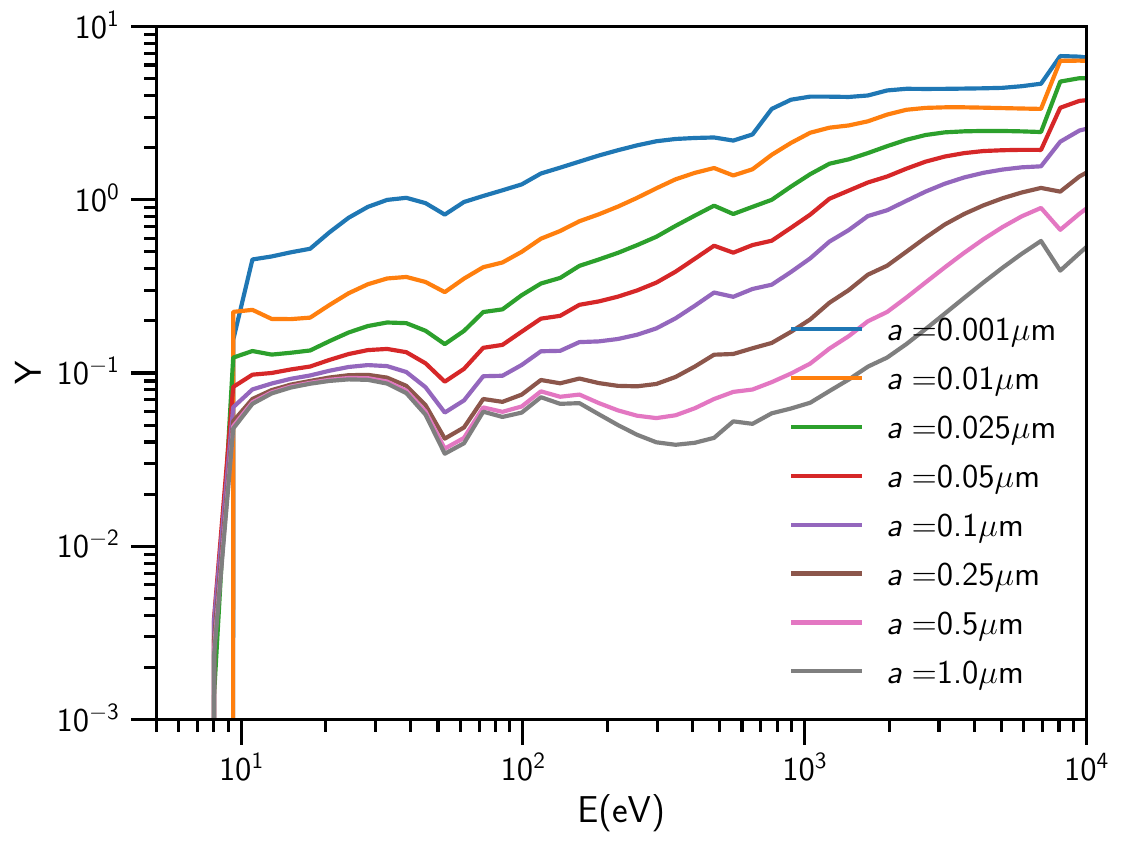}
\caption{Photoelectric yield of neutral silicate grains as a function of photon energy for the different grain sizes. The yield is larger for smaller grains.}
\label{fig:Y}
\end{figure}

Figure \ref{fig:Y} shows the total photoelectric yield as a function of photon energy for a neutral silicate grain of different sizes. For the range of interstellar UV radiation, $W<h\nu<13.6\eV$, the photoelectric yield is almost constant. The yield increases with the photon energy for $h\nu>40$ eV due to the emission of primary, Auger, and secondary electrons.

For convenience, we parameterize the photoelectric yield for UV photons of $13.6\eV>h\nu>W$ using the numerical calculations of the photoelectric yield shown in Figure \ref{fig:Y}, 
\bea 
Y(a) = \left\{
\begin{array}{l l}   
    0.7 ~ ~  {\rm ~ for~ } a \le 0.001\mum   \\
    0.5 ~ ~  {\rm ~ for~ } 0.001\mum <a \le 0.005\mum   \\
    0.3 ~ ~  {\rm ~ for~ } 0.005\mum <a < 0.01\mum   \\
    0.2 ~ ~  {\rm ~ for~ } 0.01\mum <a < 0.05\mum\\
    0.1    ~ ~ ~ ~  {\rm ~ for~} 0.05< a <1 \mum \\
\end{array}\right.,
\label{eq:Ype}
\ena
and $Y=0$ for $h\nu<W$. 

Note that for X-rays (of energy above $100$ eV), the photoelectric yield increases due to the contribution of Auger and secondary electrons (see Figure \ref{fig:Y} and \citealt{Hoang.2015}). However, the energy density of diffuse interstellar X-ray is much lower than UV photons, so it is ignored in this section. 



\subsubsection{SPE photoemission rate and yield}
The total rate of photoelectron emission per H (electron/s/H) is then
\bea
\Gamma_{\rm pe}&=&\int_{a_{\rm min}}^{a_{\rm max}}J_{\rm pe}(a)\frac{n_{\H}^{-1}dn}{da} da\nonumber\\
&=& \int_{\nu_{\rm pet}}^{\infty}  \frac{cu_{\nu}}{h\nu} d\nu\int_{a_{\rm min}}^{a_{\rm max}}Y(a,\nu)Q_{\rm abs}\pi a^{2}\frac{n_{\H}^{-1}dn}{da} da
,\label{eq:Gamma_pe}
\ena
where $n_{\H}$ is the hydrogen density in the interstellar gas, $dn/da = Ca^{-3.5}$ is the grain size distribution with the lower and upper cutoff $a_{\rm min}, a_{\rm max}$, respectively \citep{Mathis.1977}. The constant $C$ is determined by the mass ratio of dust to the gas, which is chosen to be $1:100$ for the typical ISM.


The emission rate of spin-polarized electrons per H is calculated from aligned grains, which reads
\bea
\Gamma_{\rm pe}^{\rm SPE}&=&\int_{a_{\rm min}}^{a_{\rm max}} f_{\rm align}(a) J_{\rm pe}(a)\frac{dn}{n_{\H}da} da,\nonumber\\
&=& \int_{\nu_{\rm pet}}^{\infty}  \frac{cu_{\nu}}{h\nu} d\nu\int_{a_{\rm min}}^{a_{\rm max}}f_{\rm align}(a)Y(a,\nu)Q_{\rm abs}\pi a^{2}\frac{n_{\H}^{-1}dn}{da} da,~~~\label{eq:Gamma_SPE}
\ena
where $f_{\rm align}$ accounts for the alignment degree as a function of the grain size (see Eq. \ref{eq:falign}).

The fraction of spin-polarized photoelectrons to the total photoelectrons is
\bea
f_{\rm SPE}=\frac{\Gamma_{\rm pe}^{\rm SPE}}{\Gamma_{\rm pe}}.\label{eq:f_SPE}
\ena

We are also interested in the yield of SPE emission, defined by the ratio of SPE emission rate to UV irradiation rate:
\bea
Y_{\rm SPE}=\frac{\Gamma_{\rm pe}^{\rm SPE}}{\Sigma_{d}cn_{\rm UV}}=\frac{\Gamma_{\rm pe}^{\rm SPE}}{\Sigma_{d}\int d\nu(cu_{\nu}/h\nu)},
\label{eq:Y_SPE}
\ena
where $\Sigma_{d}$ is the total dust surface area per H given by
\bea
\Sigma_{d}&=&\int_{a_{\rm min}}^{a_{\rm max}} da \pi a^{2} (n_{\rm H}^{-1}dn/da)=C\left(a_{\rm min}^{-0.5}-a_{\rm max}^{0.5}\right)\nonumber\\
&=& 10^{-21}\left(\frac{C}{10^{-25}\cm^{2.5}}\right)\left(\frac{a_{\rm min}}{0.001\mum}\right)^{-0.5}\cm^{2}\H^{-1}.~~~\label{eq:Sigma_d}
\ena

For the diffuse interstellar UV, using Equation (\ref{eq:u_MMP}) we estimate the density of UV photons to be $n_{\rm UV}\sim 0.0054G_{\rm UV}\cm^{-3}$. Therefore, the SPE emission yield (per incident UV photons) is 
\bea
Y_{\rm SPE}\simeq 0.067 \left(\frac{\Gamma_{\rm SPE}}{10^{-14}\s^{-1}\H^{-1}}\right)\left(\frac{10^{-21}\cm^{2}\H^{-1}}{\Sigma_{d}}\right).\label{eq:Y_SPE}
\ena

Equation (\ref{eq:Y_SPE}) implies $Y_{\rm SPE}\sim 7\%$ for the UV photoelectric yield of $Y\sim 0.3$, i.e., 7 SPEs are produced for every 100 incident UV photons. Extreme UV or X-ray would produce higher SPE yield due to higher photoelectric yield (see Figure \ref{fig:Y}).

\subsection{Anisotropy and Polarization of SPE Photoemission}
For an isotropic radiation field, the emission of SPEs is still anisotropic due to the alignment of grains with the magnetic field that results in the differential cross-section with the ejection along and in the direction perpendicular to the magnetic field (see Figure \ref{fig:Barnett}). Assuming the isotropic photoelectric yield, the anisotropy of SPEs is defined by
\bea
\gamma_{\rm SPE}=\frac{\Gamma_{\|}^{\rm SPE}-\Gamma_{\perp}^{\rm SPE}}{\Gamma_{\|}^{\rm SPE}+\Gamma_{\perp}^{\rm SPE}},
\ena
where $\Gamma_{\|,\perp}^{\rm SPE}$ are the SPE emission rate along and in the direction perpendicular to the magnetic field.

For the cylindrical grain shape of radius $r$ and height $h$, we get
\bea
\gamma_{\rm SPE}=\frac{2\pi r^{2}-2\pi rh}{2\pi r^{2}+2\pi rh}=\frac{1-s}{1+s}
\ena
which implies $\gamma_{\rm SPE}=1/3$ for $s=h/r=1/2$ and the anisotropy increases with grain elongation $s$.

One interesting feature of SPEs from aligned grains is the difference in the polarization state of SPEs with respect to the direction of electron emission. Indeed, SPEs emitted along the alignment axis have spins directed along the direction of motion, and those emitted in the direction perpendicular to the alignment axis have spins perpendicular to the direction of motion. 

\section{Numerical results}\label{sec:results}

\subsection{Dependence of SPE emission on grain alignment}
The efficiency of grain alignment is a key parameter for the SPE photoemission. According to the RAT alignment theory, the minimum size of grain alignment $a_{\rm align}$ depends on several physical parameters, including the gas density and radiation field (see \citealt{Hoang.2021}), whereas the maximum alignment efficiency $f_{\rm max}$ depends on the magnetic susceptibility and the gas density (see \citealt{HoangBao.2023}). To get insights into the effect of grain alignment on SPEs, here we calculate SPE photoemission for different values of $a_{\rm align}$ and $f_{\rm max}$ instead of using their values predicted by the RAT theory.

\begin{figure*}
\includegraphics[width=0.33\textwidth]{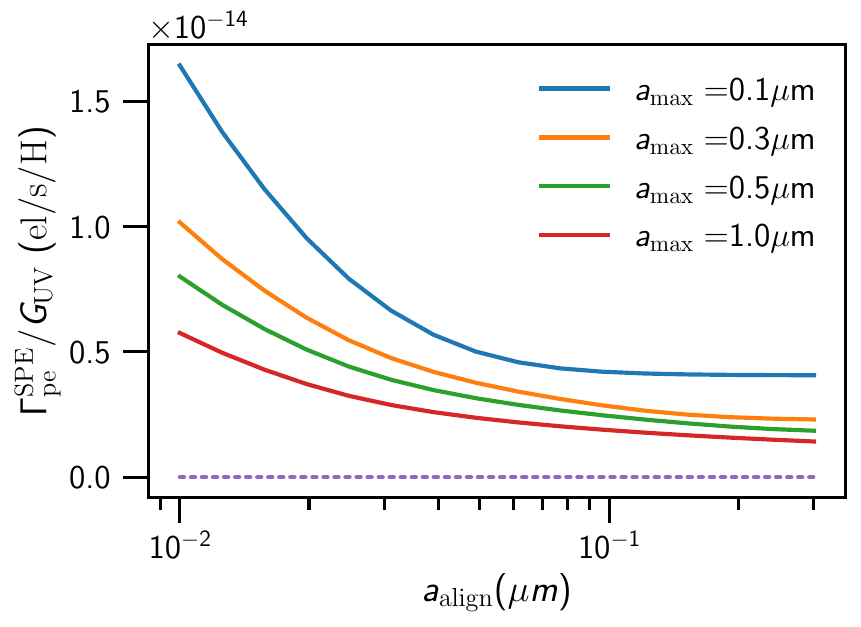}
\includegraphics[width=0.33\textwidth]{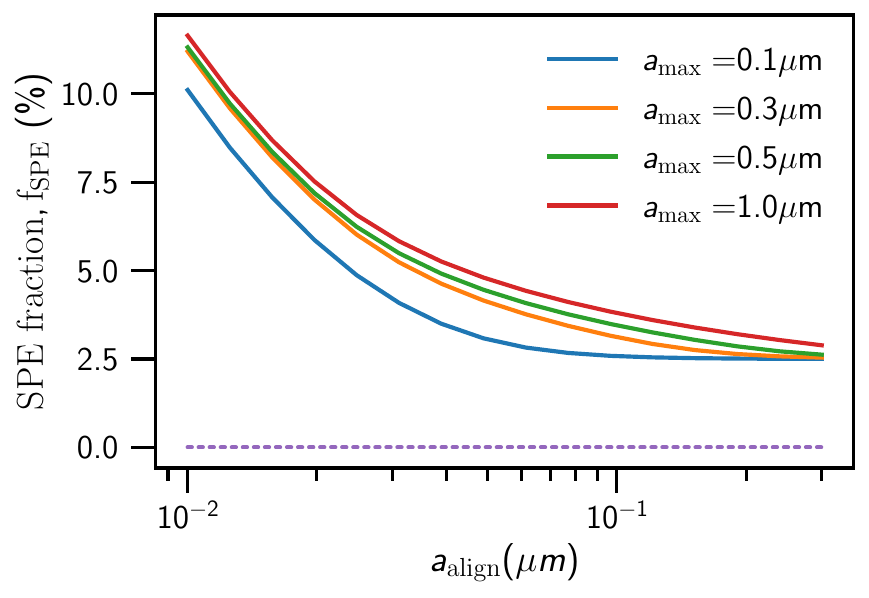}
\includegraphics[width=0.33\textwidth]{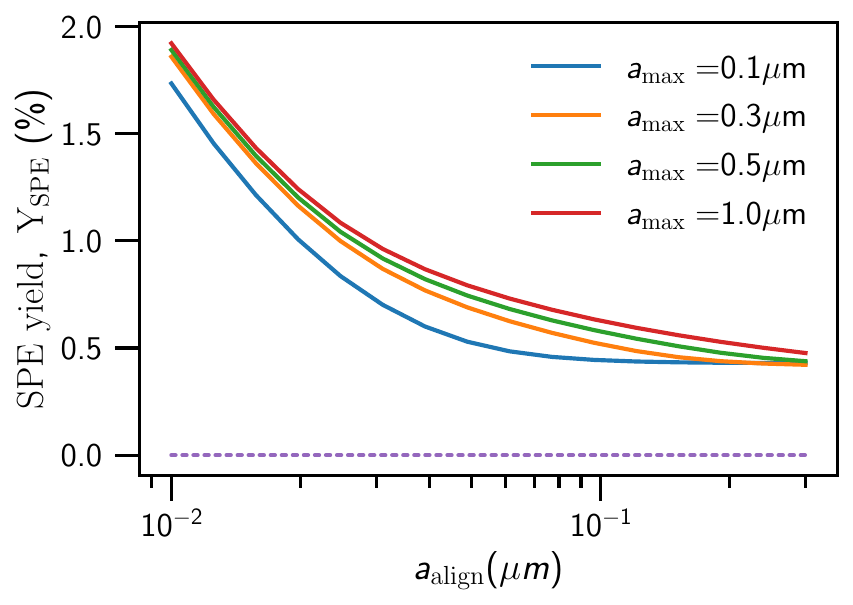}
\includegraphics[width=0.33\textwidth]{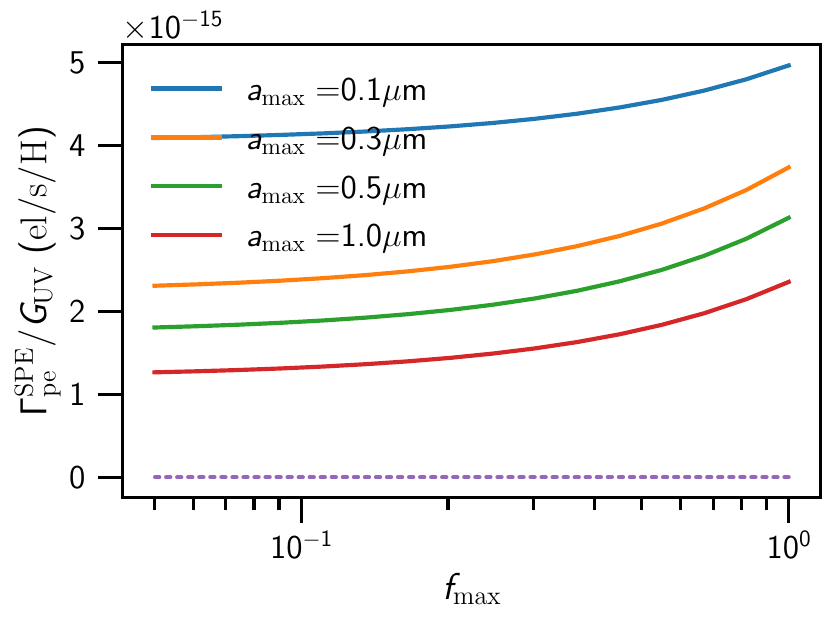}
\includegraphics[width=0.33\textwidth]{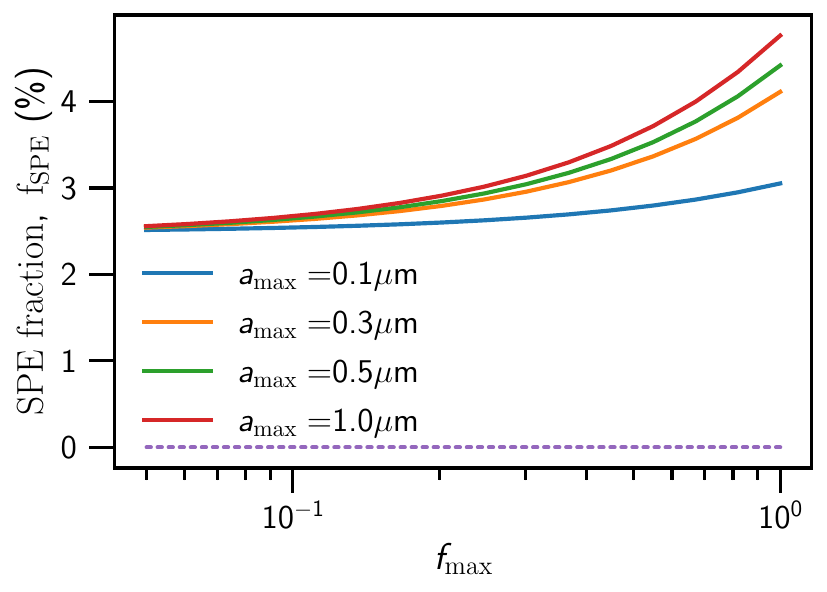}
\includegraphics[width=0.33\textwidth]{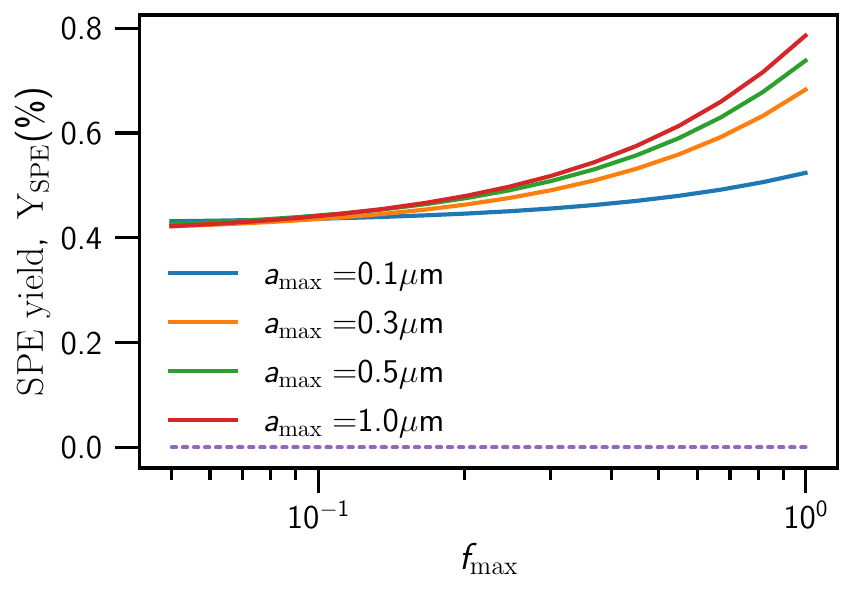}
\includegraphics[width=0.33\textwidth]{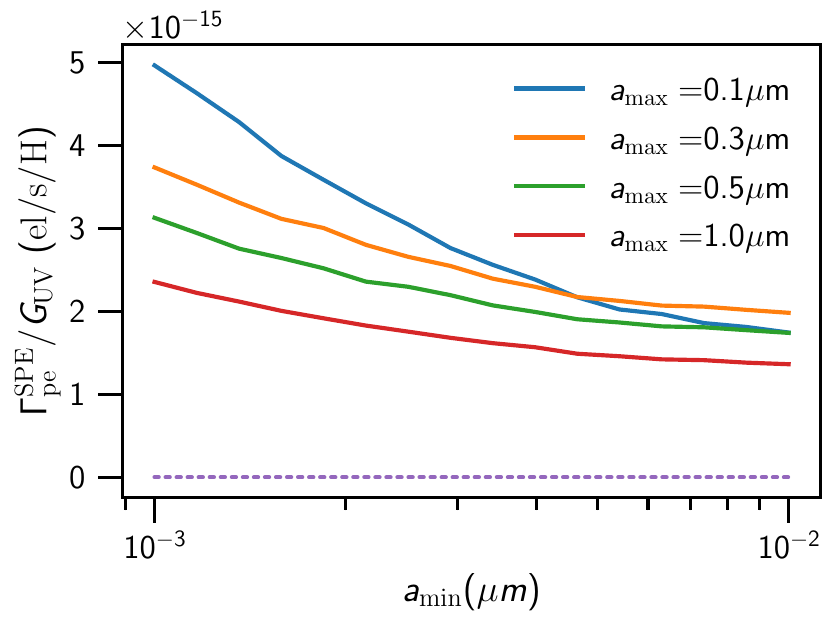}
\includegraphics[width=0.33\textwidth]{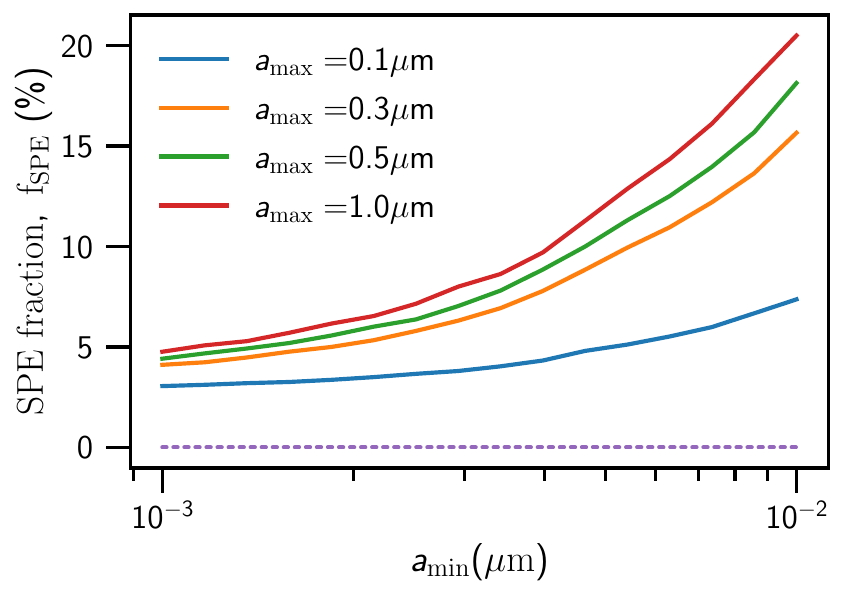}
\includegraphics[width=0.33\textwidth]{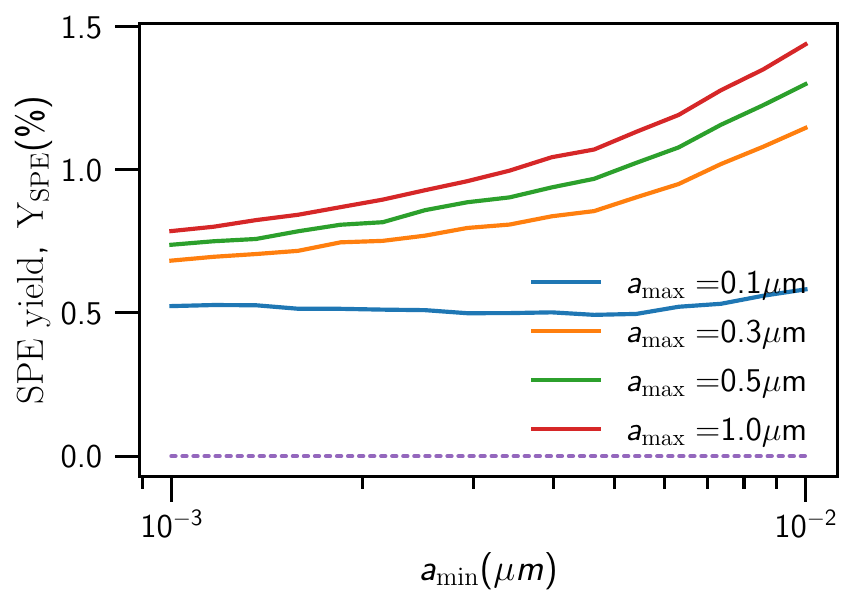}
\caption{Top row: photoemission rate of SPE per H (left), fraction of SPE to the total photoelectrons as a function of the grain alignment size (middle), and SPE photoemission yield (SPE per UV photon) (right). The dashed horizontal line shows the results produced by the smallest grains of $a<a_{\rm align}$. Perfect alignment of large grains of $a>a_{\rm align}$ with $f_{\rm max}=1$ are considered.
Middle row: Same as the top row but for the variation of the photoemission rate for the different maximum alignment degrees. The ratio of SPE increases with grain growth and could reach $15\%$ for $a_{\rm min}=0.01\mum$.
Bottom row: Same as the top row but for the variation of the photoemission rate for the minimum size of the grains. The ratio of SPE increases with grain growth and could reach $15\%$ for $a_{\rm min}=0.01\mum$. Perfect alignment of large grains of $a>a_{\rm align}$ with $f_{\rm max}=1$ are considered.}
\label{fig:fSPE_aali_fmax}
\end{figure*}

Figure \ref{fig:fSPE_aali_fmax} (top row) shows the results for photoemission of SPEs from aligned grains as a function of the alignment size, $a_{\rm align}$, for the different maximum grain size, $a_{\rm max}$, assuming the typical ISRF and the perfect alignment of large grains with $f_{\rm max}=1$. The left, middle, and right panels show the total SPE emission rate ($\Gamma_{\rm pe}^{\rm SPE}/G_{\rm UV}$), the fraction of SPEs relative to total photoelectrons ($f_{\rm SPE}$), and the SPE yield, respectively. One can see that aligned dust can create more than 10-20 SPEs per H. Moreover, both the SPE photoemission rate and the fraction of SPEs increase rapidly with decreasing $a_{\rm align}$ due to the increase in the fraction of aligned grains. This is reasonable because small grains contribute more to the total surface area and the photoelectric yield is larger for small grains. For the typical ISRF, the alignment size predicted by RATs is $a_{\rm align}\approx 0.05\mum$ (see \citealt{Hoang.2021}), for which the fraction of SPEs is $f_{\rm SPE}\approx 5\%$.

The contribution of nanoparticles produces about $2.5\%$ of SPEs due to the weak alignment of nanoparticles. Numerical calculations by \cite{HoangLaz.2016b} showed that nanoparticles could reach the maximum alignment degree of $\sim 5\%$. However, due to their dominant surface area and higher photoelectric yield, the emission of SPEs is still contributing about $2.5-5\%$.

Figure \ref{fig:fSPE_aali_fmax}(middle row) shows similar results as the top row but as a function of the alignment efficiency $f_{\rm max}$. The SPE rate increases slightly with $f_{\rm max}$ and reaches $5\%$ for perfect alignment. For weakly aligned large grains, the emission of SPEs is only produced by nanoparticles. Interestingly, one can see that increasing the maximum grain size $a_{\rm max}$ tends to decrease the SPE emission rate due to the reduction in the surface area of the largest grains (upper panel) but increases the SPE fraction and yield (middle and lower panels).

\subsection{Dependence of SPE emission on grain growth}

Grain growth due to gas accretion and grain-grain collision is expected to occur in dense clouds (see e.g., \citealt{Hoang.2022}). The grain growth increases the surface area of large grains but reduces that of the smallest grains, which affects photoemission rates. To study the effect of grain growth on SPEs, we calculate the SPE photoemission for the different $a_{\rm min}$.

Figure \ref{fig:fSPE_aali_fmax}(bottom row) shows similar results as the top row but as a function of the lower cutoff of the grain size distribution. The minimum size accounts for the effect of grain growth in molecular clouds, especially in star-forming regions. The SPE rate increases slightly with $a_{\rm min}$, but the ratio of SPE increases with $a_{\rm min}$ due to the reduction in the photoemission from small and unaligned grains. In particular, the grain growth that reduces the smallest grains of $a<0.01\mum$ could enhance the fraction of SPEs to $10\%$. This situation is important for photodissociation and HII regions irradiated by UV radiation from young stellar objects and massive star-forming regions. Same as the top and middle rows, increasing the maximum grain size $a_{\rm max}$ tends to decrease the SPE emission rate (top panel) but increases the SPE fraction and yield (middle and bottom panels).

\section{The Role of Aligned Magnetic Grains in Chiral Asymmetry}\label{sec:role_chiral}
The role of aligned magnetic grains in chiral asymmetry is exhibited through the three effects. First, magnetically aligned grains are the source of SPEs due to the photoelectric effect induced by interstellar UV photons. Second, the resulting SPEs cause chiral asymmetry through reduction chemical reactions with chiral molecules in the grain surface (i.e., CDRC effect). Third, magnetic grains containing aligned spins become a magnetic agent that enables the spin-selective adsorption (CISA) of chiral molecules due to the spin-spin effect. Below, we discuss these effects in more detail.

\subsection{Effects of SPEs on Chiral Molecules via the CISS effect}

The CISS effect was first discovered by \cite{Ray.1999} where the authors found that the transmission of electrons through chiral molecules is spin-dependent. The later experiment by \cite{Gohler.2011} showed that transmitted electrons through doubled-stranded (chiral) DNA are spin-polarized, and photoelectrons are spin-polarized even when they are produced by unpolarized light. 

Although a quantitative theory of the CISS effect is not yet available (see \citealt{Evers.2022} for a review), the basic physics can be summarized as follows. Chiral molecules exhibit an helical electric field $\bE_{\rm helix}$ due to helical charge distribution (i.e., asymmetric charge distribution) caused by the molecule configuration \citep{Naaman.2012,Gutierrez.2012}. As the electron moves through a chiral molecule with velocity $\bv$ in an electric field $\bE_{\rm helix}$, in the co-moving electron's frame, there exists an effective magnetic field of $\bB_{\rm eff}=-[\bv/c\times \bE_{\rm helix}]$. The interaction between the electron magnetic moment and the effective magnetic field induces a potential energy
given by (\citealt{Manchon.2015}, see also Appendix \ref{apdx:CISS})
\bea
V_{\rm SO}=-\bmu_{e}.\bB_{\rm eff}=\frac{g_{e}\mu_{B}}{2\hbar} \bS.[-\bv/c\times \bE_{\rm helix}],
\ena
where $\bmu=\gamma_{e}\bS=-(g_{e}\mu_{B}/\hbar) \bS$ with $\bS$ the electron spin. This potential determines the coupling between the electron spin and its linear momentum ($m_{e}\bv$), known as the Rashba spin-orbit effect. The Rashba SO effect induces the Zeeman splitting of the electron energy level into three levels with the energy difference is $2V_{\rm SO}$. 

For each interaction of a SPE with a chiral molecule, there are two possible states of the electron spin (up and down) and two states of the linear momentum (moving upward (+) and downward (-)). The four energy states are described by $\ket{+,\uparrow}$, $\ket{+,\downarrow}$, $\ket{-,\uparrow}$, $\ket{-,\downarrow}$. The direction of the chiral electric field is defined such that an electron moving upward through a right-handed (dextral or D-) molecule has a positive effective magnetic field. Therefore, for D-molecules, one has $V_{\rm SO}>0$ for the states $\ket{+,\uparrow}$ and $\ket{-,\downarrow}$, and $V_{\rm SO}<0$ for $\ket{+,\downarrow}$ and $\ket{-,\uparrow}$. Similarly, for L-molecules, $V_{\rm SO}>0$ for the states $\ket{+,\downarrow}$ and $\ket{-,\uparrow}$, and $V_{\rm SO}<0$ for $\ket{+,\uparrow}$ and $\ket{-,\downarrow}$. 

Let $\ket{D}=\{\ket{+,\uparrow},\ket{-,\downarrow}\}$ and $\ket{L}=\{\ket{+,\downarrow},\ket{-,\uparrow}\}$ denote the degenerate double helical states of the D-electron and L-electron respectively. Thus, D-molecules prefer to interact with $\ket{L}$ electrons and vice versa.


The probability for spin flipping between two $\ket{D}$ and $\ket{L}$ states is 
\bea
P_{\rm flip}=\exp\left(\frac{-2V_{\rm SO}}{kT}\right),\label{eq:Pflip}
\ena
where $T$ is the ambient temperature \citep{Naaman.2012}.

The effective values of $V_{\rm SO}$ are measured to be $20-100$ meV for chiral molecules \citep{Naaman.2012}, and larger molecules could reach $500$ meV (see \citealt{Ozturk.2022}).\footnote{The value $V_{\rm SO}$ is experimentally measured, but the calculated SO energy is much lower. Therefore, the theory of the CISS effect is still catching up with the experiments, and it is suggested that an additional interaction such as spin-spin exchange may enhance the potential of chiral molecule-electron spins.} Let $T_{\rm cri}$ be the critical temperature above which thermal collisions can randomize the Zeeman energy levels, i.e., destroying the SO effect. Then, 
\bea
T_{\rm cri}=\frac{V_{\rm SO}}{k}\approx 100\K \left(\frac{V_{\rm SO}}{100 {\rm meV}}\right).\label{eq:Tcri}
\ena

In the ISM and cold regions, the gas and dust temperature is typically of $T<100\K$. Therefore, thermal collisions are insufficient to destablize the the Zeeman levels because because $p_{\rm flip}=e^{-T_{\rm cri}/T}\ll 1$.

\subsection{Grain surface CISS-driven reduction chemistry and enantiomer excess}
Here, we discuss a specific scenario where SPEs may affect the chiral asymmetry for chemical reactions in the ice mantle of dust grains. Organic molecules are known to most likely form in the ice mantle of dust grains (e.g.,\citealt{Herbst:2009go}). Therefore, the interaction of SPEs (either from the grain core or from the ambient medium) with chiral molecules in the ice mantle would enrich the chiral asymmetry.

Following the CISS-driven reduction chemistry (CDRC; \citealt{Ozturk.2022}), we consider two reduction reactions for a grain mantle consisting of chiral molecules $A$ of L- and D-states:
\bea
(L/D)-A +e^{-} \rightarrow (L/D)- B^{-},\label{eq:LD_e}
\ena
where $e^{-}$ is the electron. 

If the electron is unpolarized, the rate of reactions is identical because the interaction between L-A and D-A molecules with $e^{-}$ are identical. However, for SPEs, the spin-orbit coupling induces the favor of the interaction of $L$-A with $D-$electron and disfavors of the $L-$A with $L-$ electron due to the energy barrier of $2V_{\rm SO}$, resulting in the asymmetry of the reduction reactions. 

Let us consider the reduction of a chiral molecule $A$ induced by a right-handed (D) electron. The chemical reduction on the grain mantle of temperature $T_{\rm grain}$ can be described by the following reactions \citep{Ozturk.2022}
\bea
D-A + D-e^{-} &&\rightarrow D-B^{-}: k_{D},\\
L-A + D-e^{-} &&\rightarrow L-B^{-}: k_{L},\label{eq:CDRC}
\ena
and the reaction rate $k_{L}$ is greater than $k_{D}$ as given by
\bea
\frac{k_{L}}{k_{D}}=\exp\left(\frac{2V_{\rm SO}}{kT_{\rm grain}}\right).\label{eq:kL_kD}
\ena

We can quantify the enantiomer excess of the resulting chiral molecule $B$ as follows:
\bea
ee_{B} = \frac{L-D}{L+D}=\frac{k_{L}-k_{D}}{k_{L}+k_{D}}=\frac{k_{L}/k_{D}-1}{k_{L}/k_{D}+1},\label{eq:ee}
\ena
which yields $ee=80\%$ even for $k_{L}/k_{D}=10$.

\begin{figure}
\includegraphics[width=0.5\textwidth]{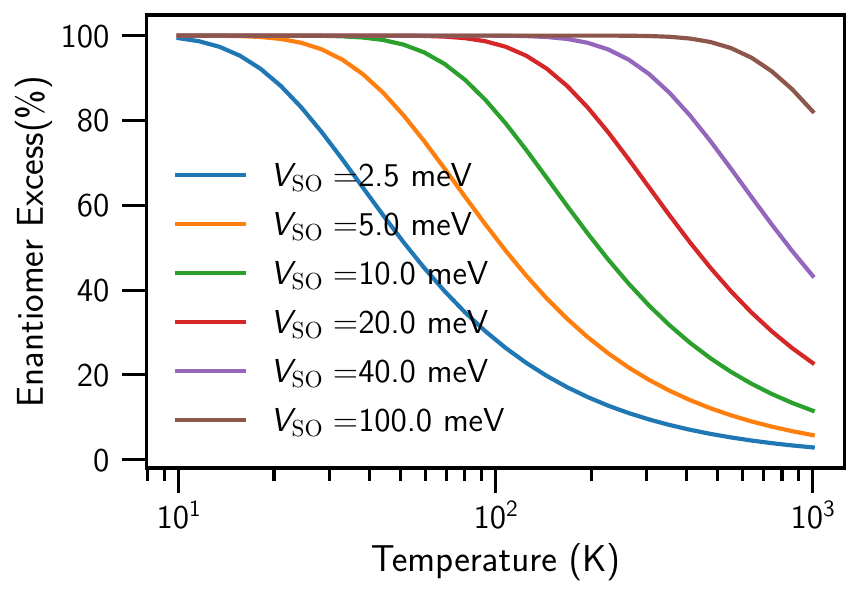}
\caption{Enantiomer excess due to CISS-driven reduction chemistry (CDRC) on the grain surface by SPEs as a function of the grain temperature for the different magnitudes of spin-orbit coupling ($V_{\rm SO}$).}
\label{fig:EE_SPE}
\end{figure}

Equation (\ref{eq:ee}) gives the enantiomer excess for chiral molecules by CDRC. Because of the lack of detection of chiral molecules on the grain ice mantle and the unknown value of $V_{\rm SO}$ of those molecules, we attempted to provide a rough estimate by considering the typical $T_{gas}$ of astrophysical environments and varying the value $V_{\rm SO}$. The numerical results are shown in Figure \ref{fig:EE_SPE} for the enantiomer excess due to CISS-driven reduction chemistry on the grain surface by SPEs as a function of the grain temperature for the different magnitudes of spin-orbit coupling. For the ISM and star-forming regions with $T_{\rm grain}<100\K$, the CDRC mostly occurs in the grain icy mantle, and the enantiomer excess could reach above $20\%$ for the considered range of $V_{\rm SO}$. At high temperatures of $T_{\rm grain}>200\K$ when the ice mantle is thermally sublimated, the CDRC can occur only in the gas phase. Because the gas temperature $T_{\rm gas}\gtrsim T_{\rm grain}$, the enantiomer excess by the gas-phase CDRC is reduced significantly.

Note that the CDRC induced by SPEs can occur on the grain mantle of all grains, regardless of their alignment. The effect can also be involved directly with interstellar big organic molecules or polycyclic aromatic hydrocarbons. Therefore, organic compounds would be most sensitive to SPEs, resulting in their chiral asymmetry.

\subsection{Grain Surface Chiral-Induced Spin-Selective Adsorption}
Experimental studies (e.g., \citealt{Banerjee-Ghosh.2018}) show that magnetic substrates can act as chiral agents, i.e., they interact stronger with one enantiomer of chiral molecules than the other due to electron dipole-dipole (i.e., spin-spin) interaction, a mechanism termed chiral-induced spin-selective adsorption (CISA). \cite{Banerjee-Ghosh.2018} first demonstrated the CISA effect for DNA and L-cysteine on a ferromagnetic thin film, and \cite{Ozturk.2023b} found the effect for RAO-a precursor of RNA on a magnetite ($F_{2}O_{3}$) surface.

\begin{figure}
\includegraphics[width=0.5\textwidth]{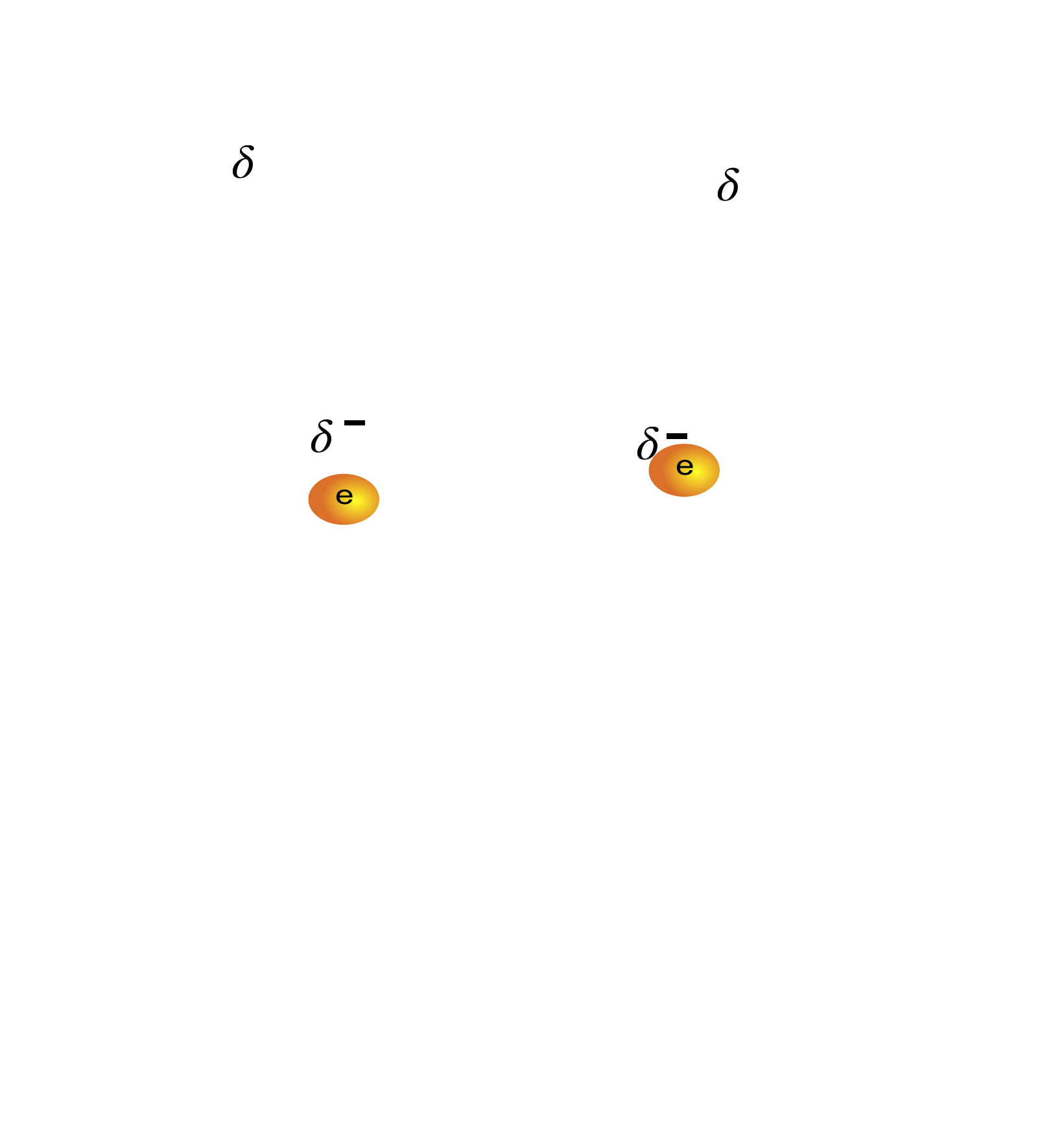}
\caption{Illustration of the chiral-induced spin-selective adsorption effect on an aligned grain with electron spinup (black arrows). When a chiral molecule (i.e., L- or D-molecule) approaches the grain surface, the charge redistribution process causes the dominance of D-electrons from L-molecule and L-electrons from D-molecule at the pole of the molecule (marked by $\sigma^{-}$ sign) near the grain surface, due to the CISS effect. The exchange interaction between the grain electron spin (black arrow) and the molecular electron spin (thick black arrow) favors the D-molecules with anti-parallel spins (i.e., attractive) and disfavor L-molecules of parallel spins (i.e., repulsive), following the Pauli exclusion principle. Thus, the rate of adsorption for D-molecules to the grain is higher than L-molecules.}
\label{fig:CISA}
\end{figure}

Here, we consider the spinning grains aligned with the ambient magnetic field to act as a magnetic surface that facilitates enantioenrichment of chiral molecules in the ISM. The schematic illustration is shown in Figure \ref{fig:CISA}.

The physics of the CISA effect can be briefly described as follows. When a chiral molecule approaches the aligned grain (a surface), the charge density in the molecule redistributes due to the effect of the electric field produced by the surface (i.e., the surface has an electric dipole, van der Waals force) so that electrons are displaced in the opposite direction of the induced electric field. The process of charge redistribution is caused by the motion of electrons across the chiral molecule, which is spin-dependent due to the CISS effect, and gives rise to a transient spin-polarization (see e.g., \citealt{Naaman.2018} for a review). 

Let's consider a L-molecule and a D-molecule approaching a grain surface. The redistribution is spin-dependent such that L-electrons are easily moved along the D-molecule whereas D-electrons are disfavored. As a result, D-electrons from the L-molecule and L-electrons of the D-molecule dominantly distribute near the grain surface. Subsequently, the interaction between L-electrons and D-electrons and D-electrons from the grain surface. Due to the Pauli exclusion principle, the exchange interaction between anti-parallel spins is favored while parallel spins are disfavored, which results in the dominance of D-molecules (see Fig. \ref{fig:CISA}). Similarly, the grain with an opposite aligned direction accumulates L-molecules.

\section{Discussion}\label{sec:discuss}
\subsection{SPE emission from aligned grains by UV, X-rays, and cosmic rays}
In Section \ref{sec:results}, we have discussed the emission of SPEs from aligned grains by the diffuse interstellar UV radiation. Here, we consider other sources of SPEs in astrophysical environments.

Massive stars and young stellar objects (YSOs) are the strong source of UV photons. In these environments, grains are efficiently aligned with the magnetic field by the MRAT mechanism \citep{Hoangetal.2022,Giang.2023}. Thus, we expect that SPEs are abundant due to enhanced SPE emission yield resulting from grain growth (see Figure \ref{fig:fSPE_aali_fmax}, bottom row). Moreover, if the UV light is circularly polarized, the amount of left- and right-SPEs from aligned grains would be different. Ly-$\alpha$ photons resulting from the recombination of electrons and protons in HII regions caused by massive stars, active galactic nuclei, or accretion disks around black holes can also be an important UV source that triggers SPEs from aligned grains.

Supernova explosions, gamma-ray bursts, and active galactic nuclei are the most powerful sources of energetic photons from EUV and X-rays in the Universe. As shown in Figure \ref{fig:Y}, the photoelectric yield of EUV to X-rays is significantly higher than that of UV photons. As a result, the SPE yield emitted from aligned grains is significantly enhanced (see Eq. \ref{eq:Y_SPE}). The enhanced flux of SPEs could be the source of chiral asymmetry and cause homochirality.

Finally, cosmic rays (CRs) can eject SPEs from aligned grains via collisional ionization \citep{Hoang.2015} and CR-induced UV and X-ray radiation can also create SPEs via the photoelectric effect on aligned grains. Since CRs can penetrate deep into dense molecular clouds and protostellar disks, CR-induced SPEs could be the main driving force for inducing chiral asymmetry of complex molecules that are formed on icy grain mantles in cold and dense clouds, protostellar cores, and planet-forming disks. 


\subsection{Aligned magnetic grains as a key agent of interstellar chiral asymmetry} 
The homochirality of biological molecules is the unique feature of life on Earth. However, the origins of such homochirality remain much debated (see \citealt{Sparks.2015,Brandenburg.2020} for reviews). Moreover, to understand whether there exists life elsewhere in the Universe, it is first to understand whether chiral molecules could be formed in the Universe and exhibit chiral asymmetry.

Complex organic molecules (COMs, e.g., C$H_{3}$OH and C$_{2}$H$_{5}$OH), the building blocks of biological molecules, are believed to first form in the ice mantle of dust grains from simple molecules such as H$_{2}0$, CO, HCN, and NH$_{3}$ \citep{Herbst:2009go}. They subsequently desorb from the grain mantle due to thermal sublimation in star-forming regions, such as hot cores around massive protostars or hot corinos around low-mass protostars \citep{Herbst:2009go,2018IAUS..332....3V}. The first gas-phase interstellar glycine were discovered in the Galactic Center (GC) and star-forming regions OMC-1 and W52, which are likely formed in icy grain mantles and desorbed by thermal sublimation \citep{Kuan.2003}. Later, \cite{McGuire.2016} detected the first chiral amino acid (propylene oxide) toward the GC by using the rotational spectroscopy observed with single-dish radio Green Bank Telescope (GBT). Recently, gas-phase ethanolamine (a precursor of phospholipids, \cite{Rivilla.2022}) and glycine isomer \citep{Rivilla.2023} were detected toward the GC using the rotational spectroscopy using Yerkes 40 m and IRAM 30 m radio telescopes.

Experimental studies by \cite{Bernstein.2002} and \cite{Caro.2002} independently demonstrated that amino acids (including glycine, alanine, and serine) could be formed naturally from UV photolysis of the analogs of interstellar ice (consisting of H$_2$O, HCN, NH$_{3}$, CH$_{3}$OH). However, such amino acids are racemic (i.e., comprising of equal numbers of enantiomers). Numerous experiments have also demonstrated that irradiation of low-energy electrons onto the ice mantles analogous to interstellar ice could trigger chemical reactions in similar ways as irradiation of UV photons \cite{Boamah.2014,Boyer.2016,Sullivan.2016,Kipfer.2024}. In particular, the experiment by \cite{Esmaili.2018} showed that glycine can be formed by irradiation of low-energy electrons on interstellar analog of CO$_{2}$-CH$_{4}$-NH$_{3}$ ice (see \cite{Arumainayagam.2019} for a review).
 
Here, we suggest that magnetic grains aligned with the ambient magnetic field could be a key agent for the symmetry breaking of chiral molecules. First, aligned grains produce low-energy SPEs due to the photoelectric effect caused by interstellar UV photons. Such SPEs interact with chiral molecules on the aligned grain ice mantle, causing the chiral asymmetry due to CISS-driven reduction chemistry (CDRC) effect \citep{Ozturk.2022}. For the ISM and star-forming regions with the grain temperature of $T_{\rm grain}<100\K$, our calculations shown in Figure \ref{fig:EE_SPE} suggest that the CDRC can produce an enantiomer excess above $20\%$ for the possible range of spin-orbing coupling potential $V_{\rm SO}=1-100$ meV (see e.g., \citealt{Naaman.2012}). At high temperatures of $T_{\rm grain}>200\K$ when the ice mantle is thermally sublimated, the CDRC can occur only in the gas phase and the enantiomer excess is reduced significantly.

In addition, aligned grains containing aligned electron spins could act as the magnetic surface that induces the chiral-selective adsorption due to exchange interaction \citep{Rosenberg.2008,Banerjee-Ghosh.2018,Ozturk.2023b}, leading to the accumulation of a preferred enantiomer of chiral molecules than the other and chiral asymmetry (see Section \ref{sec:role_chiral}). Finally, aligned grains can act as a chiral agent which helps form more complex molecules of similar chirality and increase chiral asymmetry thanks to magnetic dipole-dipole interaction (see Appendix \ref{apdx:grain_chiral}).

\subsection{The role of interstellar magnetic fields on SPEs from dust and chiral asymmetry}
In our proposed mechanism of SPEs, magnetic fields play a crucial role in aligning dust grains via Larmor precession and magnetic relaxation \citep{HoangLaz.2016,HoangLaz.2016b}. Moreover, magnetic fields also affect the propagation of SPEs in the ISM due to the Lorentz force. For instance, one particular feature is the difference in polarization of SPEs. For SPEs emitted along the alignment axis, they have spin along the direction of motion and those emitted perpendicular to the alignment axis have spin perpendicular to the direction of motion. We call longitudinal and transverse spin polarization. Transverse spin SPEs would be constrained by the magnetic fields while longitudinal SPEs can move along the magnetic field line. Therefore, longitudinal SPEs may be dominant over transverse ones. It is noted that the local magnetic fields are turbulent, which causes the small-scale fluctuations of the momentum of SPEs and their spins.

Finally, large-scale galactic magnetic fields determine the orientation of dust grains on galactic scales and the net spin polarization of SPEs. Therefore, within our proposed mechanism, magnetic fields appear to be a key player in the chirality symmetry breaking and the origin of life. Interestingly, Louis Pasteur predicted that magnetic fields may play a role in inducing homochirality of the biological world.

\subsection{Disalignment in spin direction of SPEs}\label{sec:disalign}

To be important for chiral symmetry breaking, low-energy SPEs must maintain their spin orientation from the creation place within the grain during their passage to the surface and in their journey through the ISM before colliding with dust grains. Here we first study the disalignment of spins of SPEs due to elastic scattering by atomic nuclei within the grain. We then discuss the disalignment and loss of SPEs when released into the ISM due to interaction with gas species.

\subsubsection{Elastic scattering of SPEs within the dust grain}
\begin{figure}
\includegraphics[width=0.5\textwidth]{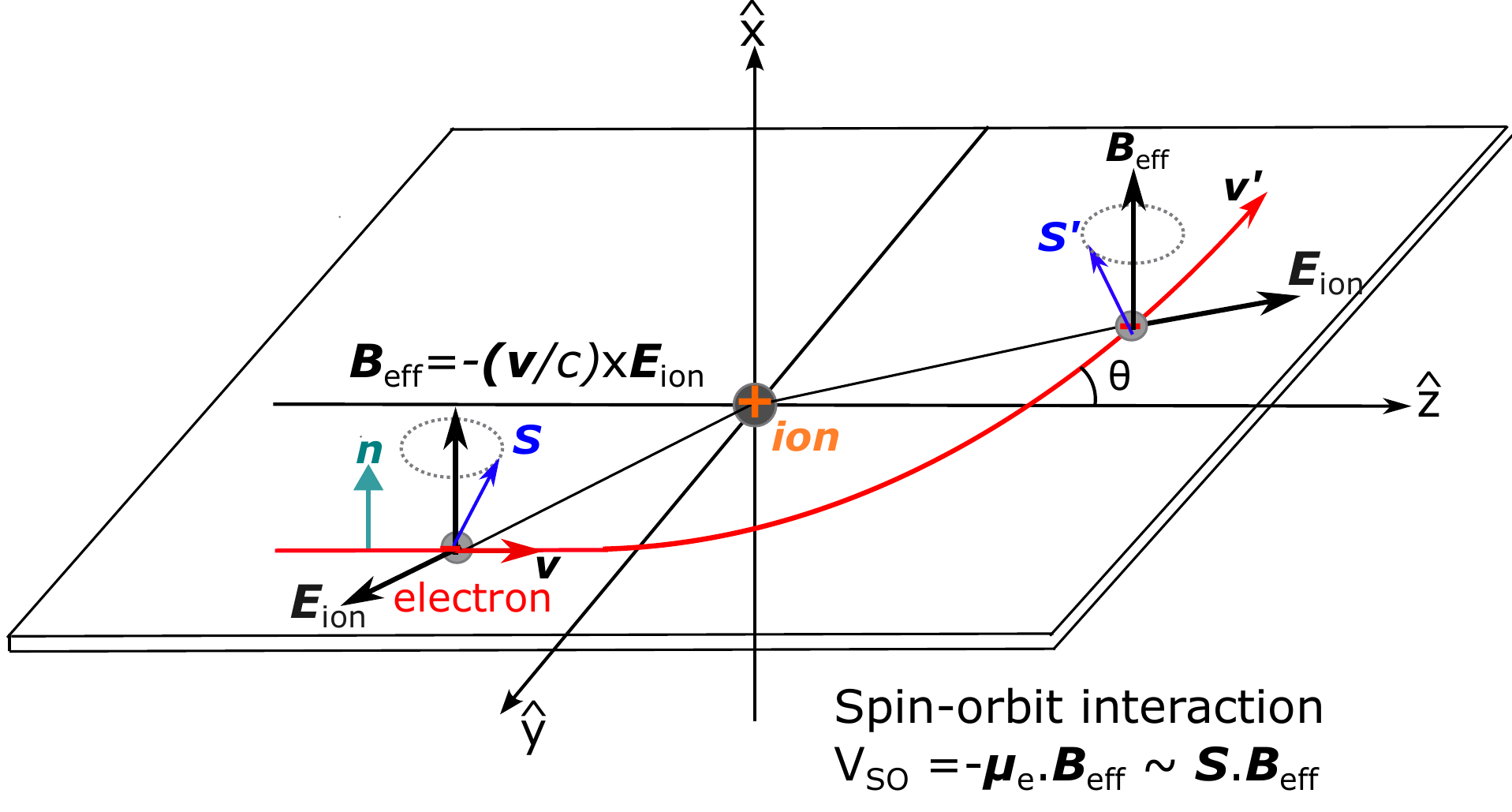}
\caption{Illustration of the polarization of electron spins by elastic scattering in a Coulomb field produced by an ion. The interaction between the electron magnetic moment $\bmu_{e}$ with the induced magnetic field $\bB_{\rm eff}$ causes the precession of the electron spin $\bS$ associated with its magnetic moment around $\bB_{\rm eff}$. The scattered electron is polarized with the spin perpendicular to the scattering plane ($\yhat\zhat$) made by $\bv$ and $\br$ with the normal vector $\bn$ if the Larmor precession is much faster than the interaction time.}
\label{fig:electron_scat}
\end{figure}

Consider an electron with aligned spin moving from the creation place to the grain surface. During this passage within the grain, SPEs can undergo elastic scattering by the grain atomic nuclei. The Coulomb interaction does not change the electron spin, but the spin-orbit (SO) coupling may disalign the electron spin. Let $\bv$ be the velocity of SPEs and $\br$ be the radius directed from the nucleus to the electron. The electric field induced by the atomic nucleus with charge number $Z$ on the electron is then $\bE=Ze\br/r^{3}$. In the electron's frame, the electron experiences an effective magnetic field given by 
\bea
\bB_{\rm eff}=-[\bv/c\times \bE]=\frac{Ze}{cr^{3}}[\br\times \bv],\label{eq:Beff}
\ena
 which is perpendicular to the scattering plane defined by $(\bv,\br)$ (see Figure \ref{fig:electron_scat}).  Therefore, if the precession time of $\mu_{e}$ (also $\bS$) around $\bB_{\rm eff}$ is shorter than the electron-nucleus interaction time, the spin of scattered electrons is on average directed along $\bB_{\rm eff}$ and is perpendicular to the scattering plane, resulting in the disalignment of electron spins by elastic scattering. If the precession time is much longer than the interaction time, the spin of SPEs is not disaligned by elastic scattering. Below, we evaluate the two characteristic timescales and study whether the spin of SPE could be disaligned by elastic scattering.

Let $b$ be the impact factor of an electron with kinetic energy $E_{e}$ that experiences the scattering in the Coulomb field of an atomic nucleus of charge $Ze$. The critical impact factor for elastic scattering can be determined by the impact factor for which the total energy of the electron-nucleus system $H_{\rm tot}= E_{e} -Ze^{2}/b=0$, which yields 
\bea
b_{\rm cri} = \frac{Ze^{2}}{E_{e}}\sim 14 \left(\frac{Z}{10}\right)\left(\frac{10\ev}{E_{e}}\right)\AA,\label{eq:brci}
\ena
which is one order of magnitude larger than the Bohr radius. Above, the typical value $Z=10$ is taken for dust grains consisting mainly Si and C. For $b<b_{\rm cri}$, the total energy is $H_{\rm tot}<0$, and the incident electron will trigger electronic excitations, corresponding to inelastic scattering.

The SO potential of the electron-nucleus interaction is 
\bea
V_{\rm SO}&=&-\bmu_{e}.\bB_{\rm eff}=\left(\frac{g_{e}\mu_{B}}{\hbar}\right) \bS.\bB_{\rm eff},\nonumber\\
&=&\left(\frac{g_{e}}{2m_{e}^{2}c^{2}}\right) \left(\frac{Ze^{2}}{r^{3}}\right)\bS.\bL,\label{eq:H_SO_electron}
\ena
where $\bL=[\br\times m_{e}\bv]$ is the electron angular momentum. 

At the critical impact factor is given by 
\bea
|V_{\rm SO}|&=&|-\bmu_{e}.\bB_{\rm eff}|=\frac{ Ze^{2}v\hbar}{2m_{e}c^{2}b_{cri}^{2}}=\frac{\hbar m_{e}v^{5}}{8c^{2}Ze^{2}}\nonumber\\
&\simeq& 8.4\times 10^{-6}\ev \left(\frac{10}{Z}\right)\left(\frac{E_{e}}{10\eV}\right)^{5/2},\label{eq:HSO_el}
\ena
which is rather small compared to the electron kinetic energy.

The characteristic timescale of the Larmor precession of the electron's spin around the magnetic field induced by the SO coupling is given by
\bea
\tau_{\rm Lar}&=&\frac{2\pi S}{\mu_{e}B_{\rm eff}}=\frac{2\pi S  }{\mu_{e}}\frac{cb_{cri}^{2}}{Zev}\\
&\simeq& 4.26\times 10^{-10}\left(\frac{Z}{10}\right)\left(\frac{E_{e}}{10\ev}\right)^{-5/2}\s,\label{eq:tau_Lar}
\ena
where the spin magnitude $S=\sqrt{s(s+1)}\hbar$ with $s=1/2$ for electrons and the effective magnetic field is estimated at the separation of $r=b_{\rm cri}$.

The interaction timescale of the electron with the nucleus is estimated by
\bea
\tau_{\rm inter}=\frac{b_{\rm cri}}{v}\simeq 7.67\times 10^{-16}\left(\frac{Z}{10}\right)\left(\frac{E_{e}}{10\ev}\right)^{-3/2}\s.\label{eq:tau_inter}
\ena

Comparing Equations (\ref{eq:tau_Lar}) and (\ref{eq:tau_inter}) one can see that the
Larmor precession is much longer than the interaction time. Therefore, the disalignment is insignificant due to single elastic scattering for low-energy SPEs. 

We can now estimate the total number of scattering events required for the complete disalignment of electron spins, denoted by $N_{\rm disali}$, as follows. Assume the scattering follows the random walk theory, the total increase in the azimuthal angle of the electron spin around the magnetic field $\bB_{\rm eff}$ after $N_{\rm sca}$ scatterings is $(\Delta \phi)^{2}=N_{\rm sca}(\delta \phi)^{2}$ with $\delta \phi =\tau_{\rm inter}/\tau_{\rm Lar}$ the azimuthal angle changed by a single scattering. Setting $\Delta \phi =2\pi$, one obtains $N_{\rm disali}=N_{\rm sca}$ with
\bea
N_{\rm disali}\sim \left(\frac{\tau_{\rm Lar}}{\tau_{\rm inter}}\right)^{2}\sim 10^{12}.
\ena

The total scattering events for a SPE within a grain of size $a$ is $N_{sca}^{\rm grain}\sim a \sigma \bar{n}_{\rm atom}\sim 10 (a/1\mum)(\sigma/10^{-18}\cm^{2})(\bar{n}_{\rm atom}/10^{23}\cm^{-3})$ with $\sigma$ is the cross-section and $\bar{n}_{\rm atom}$ is the mean number density of atoms within the dust grain. One can see that $N_{\rm disali}\gg N_{\rm sca}^{\rm grain}$, so that the spin of SPEs is not randomized when moving within the grain from their creation place to the surface.

\subsubsection{Interaction of SPEs with the ISM}
When released into the ISM, low-energy SPEs can interact with the gas species through several processes, including electron-ion elastic scattering, electron-ion capture, electron-neutral capture, and electron-dust recombination. The first process does not affect the direction of electron spins as shown in the previous section, but the two later ones result in the loss of SPEs and the disalignment of electron spins.

Consider an ionized interstellar gas of nucleon density $n_{\H}$ and ionization fraction of $x_{\rm ion}$. SPEs can be captured by proton $H^{+}$ to produce neutral hydrogens. The cross-section for e-proton capture is $\sigma_{ep}\sim 10^{-20}\cm^{2}$ for low-energy SPEs of energy $E_{e}=1eV$\footnote{The energy distribution of photoelectrons from the interstellar UV can be described by a parabolic function with the peak around $\sim 1$ eV \citep{Draine.1978}.} (see Table 4 in \citealt{Spitzer.1948}). Thus, the mean free path (mfp) of SPEs is $\bar{\lambda}_{ep}=1/(x_{\rm ion}n_{\H}\sigma_{ep})\sim 10^{18}\cm (x_{ion,-2}n_{4})^{-1}(10^{-20}\cm^{2}/\sigma_{ep})$ where $x_{-2}=x_{\rm ion}/10^{-2}$ and $n_{4}=n_{\H}/10^{4}\cm^{-3}$. 

We now estimate the mfp of SPEs for collisions with dust grains. Assuming the dust-to-gas mass ratio $f_{d/g}=0.01$ and the single-size $a$ of grains, the average density of dust grains is $n_{\rm grain}=f_{d/g}n_{\H}m_{\H}/(4\pi a^{3}\rho_{d}/3)\sim 4\times 10^{-4}(\rho_{d/g}/0.01)n_{8}a_{0.1\mum}^{-3}(\rho_{d}/3\g\cm^{-3})\cm^{-3}$ where $\rho_{d}$ is the mass density of dust grains and $a_{0.1\mum}=a/0.1\mum$. The mfp of SPEs emitted from one grain can move before being captured by another grain is $\bar{\lambda}_{edust}\sim 1/(n_{\rm grain}\pi a^{2})\sim 10^{17}n_{4}^{-1}a_{0.1\mum}^{-1}\cm$. Comparing the mfp of e-p capture to e-dust recombination yields $\lambda_{ep}/\lambda_{edust}=10x_{ion,-2}^{-1}a_{0.1\mum}$, which implies that SPEs from one dust grain will not be captured by protons before colliding with another grain for $x_{\rm ion}<10^{-2}$. This is consistent with the picture that in dense and low ionization gas, the dominant charge carrier is dust. 

SPEs can also be captured by neutral hydrogen (H) due to dipole polarization interaction to form $H^{-}$. The cross-section of the e-H capture is $\sigma_{eH}\sim 10^{-22}\cm^{2}$ for low-energy SPEs of energy $E_{e}=1eV$ (see Table 4 in \citealt{Spitzer.1948}), which is much lower than the cross-section of e-p capture. The mean free path SPEs is then $\bar{\lambda}_{eH}=1/(n_{\H}\sigma_{eH})\sim 10^{18}n_{4}^{-1}\cm$. Comparing to the mfp of e-dust capture, one obtains $\lambda_{eH}/\lambda_{edust}=10a_{0.1\mum}$, which implies that the SPEs will not be captured by neutrals before colliding with another grain.

Inelastic collisions of SPEs with ionized metals in the gas have higher cross-section than that with protons and neutral hydrogen. However, the abundance of ions in the gas is much lower, and would not cause much disalignment of electron spins.

In summary, our numerical estimates show that the spin of SPEs is not randomized by elastic scattering during their passage from the creation place to the grain surface and their journey in the ISM. For SPEs emitted from the grain surface, they will be captured by protons in ionized gas with a considerable ionization fraction of $x_{\rm ion}>10^{-2}$. However, in the gas with lower ionization fraction, SPEs will maintain their spin alignment before being captured by dust grains.

\subsection{Depolarization of spin-polarized electrons by elastic scattering}
Depolarization of SPEs can arise from three processes, including (i) elastic scattering by atomic nuclei, (ii) inelastic scattering by atomic electrons in which the incident electron flips its spin, (iii) and exchange scattering between incident electrons and atomic electrons that have opposite spins \cite{Rose.1939}. As previously shown, inelastic scattering and exchange scattering processes are much weaker than elastic scattering in the polarization/depolarization of SPEs \citep{Rose.1939}. Therefore, we will estimate the depolarization by the most efficient elastic scattering process. 

To quantify the depolarization by elastic scattering, we need to calculate the polarization of SPEs after each scattering and compare it with the initial polarization. For an atomic nuclueus of charge number $Z$, the minimum cross-section for elastic scattering of electrons of energy $E_{el}$ can be estimated as
	\bea
	\sigma_{\rm elastic}=\pi b_{cri}^{2}=6.5\times 10^{-14}\cm^{2}\left(\frac{Z}{10}\right)^{2}\left(\frac{10 \eV}{E_{el}}\right)^{2}.\label{eq:sigma_elastic}
	\ena
	
	Consider a beam of SPEs with the number of spin-up electrons $N_{\rm up}$ and spin-down $N_{\rm down}$. The initial polarization of the incident beam is given by \citep{Kessler.1969,Gay.2009}
	\bea
	P=\frac{N_{\rm up}-N_{\rm down}}{N_{\rm up}+N_{\rm down}},\label{eq:Pol}
	\ena
	where $P$ is positive for $N_{\rm up}>N_{\rm down}$ and negative $N_{\rm up}<N_{\rm down}$. Perfect polarization beam corresponds to $P=\pm 1$.
			
	In the presence of electron spin, the total interaction potential of the electron-nucleus system comprises the Coulomb potential and the SO potential term, $V_{\rm SO}$, given by Equation (\ref{eq:HSO_el}). As a result, the differential scattering cross-section of electrons with the initial polarization $\bP$ to angle $\theta$ does not only depend on the scattering angle $\theta$ but also the azimuthal angle $\psi$ \citep{Kessler.1969,Gay.1992}:
	\bea
	\frac{d\sigma(\theta,\psi)}{d\Omega} = \sigma_{0}(\theta)[1+ S(\theta)\bP.\hat{\bn}]=\sigma_{0}(\theta)[1+ S(\theta)P\cos\psi],\label{eq:Mott_cross-section}
	\ena
	where $\sigma_{0}(\theta)$ is the spin-average scattering cross-section, $S(\theta)$ is the assymetry function which describes the effect of SO coupling on the electron scattering (aka Sherman function, \citealt{Sherman.1956}), $\hat{\bn}$ is the normal vector of the scattering plane defined by $\hat{\bn}=[\bk\times \bk']$ with $\bk$ and $\bk'$ denoting the incident and scattereding direction (see Figure \ref{fig:electron_scat}). The magnitude of $S(\theta)$ depends on $V_{\rm SO}$, which is a function of the energy of the incident electrons and the atomic number of nuclei (see Eq. \ref{eq:HSO_el}).

	Following the conventional notation, we define the spin-up/spin-down vector parallel/antiparallel to the normal vector $\hat{\bn}$. We consider the case of maximum assymmetry that occurs when the polarization is perpendicular to the scattering plane, i.e., $\psi=0$. The number of spin-up and spin-down electrons scattered through a given angle $\theta$ to the left ($\psi=0$) is $dN'_{\rm up,down}=N_{\rm up,down}(d\sigma/d\Omega)_{\rm up,down}$, which becomes
	\bea
	N'_{\rm up}(\theta) = N_{\rm up}[1+S(\theta)],\\
	N'_{\rm down}(\theta) = N_{\rm down}[1-S(\theta)],\label{eq:Nup_down}
	\ena
	which shows the increase in the number of spin-up electrons and decrease of spin-down electrons. 
	
The polarization degree of the scattered beam, $P'$, is
	\bea
P'(\theta)&=&\frac{N'_{\rm up}-N'_{\rm down}}{N'_{\rm up}+N'_{\rm down}}\\
&=&
\frac{N_{\rm up}[1+S(\theta)]-N_{\rm down}[1-S(\theta)]}{N_{\rm up}+N_{\rm down} + (N_{\rm up}-N_{\rm down})S(\theta)}\nonumber\\
&=&\frac{P+S(\theta)}{1+PS(\theta)},\label{eq:P2_sca}
\ena
which follows that $P'=S(\theta)$ for initially unpolarized beam, i.e., $P=0$, and $P'=P=1$ for completely polarized beam.
The depolarization of the SPE beam after a single elastic scattering is calculated by
\bea
DP = P- P'(\theta)=\frac{S(\theta)[P^{2}-1]}{1+PS(\theta)}.\label{eq:depol}
\ena
which strongly depends on the assymmetric Sherman function and the SO coupling strength. For an initial beam of completely polarized with $P=1$, the depolarization is zero, and the polarization of SPEs does not change. 

Numerical calculations by \cite{Sherman.1956}, the value of $S$ is rather small for small scattering angles of $\theta$ and increases with $\theta$ to the peak at $\theta\sim 90^{\circ}$. This means that the depolarization is strongest for large-angle scattering and weakest for small-angle scattering \citep{Rose.1939}. 

Note that the Sherman function is related to the asymmetry parameter introduced by \cite{Mott.1929} as $S^{2}=\delta$. \cite{Mott.1929} obtained the maximum asymmetry parameter $\delta_{\rm max}$ at the right angle scattering of $\theta=90^{\circ}$ as 
\bea
\delta_{\rm max}=11.2\times \left(\frac{Z}{137}\right)^{2}\frac{\beta^{2}(1-\beta^{2})}{(2-\beta^{2})},\label{eq:delta_max}
\ena
where $\beta=v/c$ (see also \citealt{Sherman.1956}).

Therefore, one can estimate the maximum value of ther Sherman function is $S_{\rm max}=\delta_{\rm max}^{1/2}$ for low-energy electrons of $\beta\ll 1$ scattered as 
\bea
S_{\rm max}\approx 10^{-3}\left(\frac{Z}{10}\right)\left(\frac{E_{el}}{10\ev}\right)^{1/2}.\label{eq:S_max}
\ena
where the typical value $Z=10$ is taken for dust grains consisting mainly Si and C. For the interstellar gas dominated by hydrogen and helium, $Z\sim 1-2$.

Plugging this maximum value in Equation (\ref{eq:P2_sca}) one can see that the polarization after the single scattering remains the same. The depolarization from Equation (\ref{eq:depol}) is negligible. This is expected from the fact that for low-energy SPEs, the SO interaction energy $V_{\rm SO}\sim 10^{-6}\eV$, which is rather small and insufficient to significantly change the cross-section for spin-up and spin-down electrons after scattering as given in Equation (\ref{eq:Mott_cross-section}).

Finally, we consider the depolarization effect due to multiple scattering by ions in dust grains or in the ISM. \cite{Rose.1939} first showed that the depolarization of a beam of 100 keV electrons due to multiple elastic scattering in a thin foil of Au with thickness of $0.25\mum$ (comparable to the typical grain size in the ISM) is rather low, of $1-2\%$. However, for low-energy SPEs scattered by typical atoms of $Z\le 10$ in the dust or the ISM gas, we expect that multiple scattering to be much weaker than the results from \cite{Rose.1939} because the depolarization per elastic scattering is negligible. Therefore, our estimates show that the depolarization of low-energy SPEs by elastic scattering by typical atomic nuclei in dust or in the ISM is negligible. The orientation of electron spins and the polarization of the beam of SPEs emitted from the aligned grains remain well aligned.

\section{Summary}\label{sec:summary}
We proposed interstellar dust grains aligned with magnetic fields as a key photoemission source of SPEs and chiral agents for interstellar chiral asymmetry. Our main findings are summarized as follows:
\begin{enumerate}

\item Spins of electrons within dust grains containing embedded iron inclusions are aligned by the Barnett magnetic field. Magnetic grains are aligned with the ambient magnetic field due to the Larmor precession, radiative torques, and magnetic relaxation, resulting in the alignment of electron spins with the interstellar magnetic field. 

\item Photoelectric emission of electrons from aligned grains by interstellar UV radiation produces spin-polarized (spin-up or spin-down) electrons. We quantified the rate of SPE emission from aligned grains using the modern grain alignment theory and found that the rate can reach $10^{-14}G_{\rm UV}$ electron per second per H. The fraction of SPEs relative to the total photoelectrons is $\sim 10-20\%$. 

\item The yield of SPE photoemission, defined by the ratio of the rate of SPEs to incident UV photons, is found to achieve $\sim 1\%$. This implies that each SPE is produced for every 100 incident UV photons. Energetic photons like X-rays can produce a higher SPE yield due to Auger and secondary electron effects.

\item The SPEs emitted from aligned grains could play an important role in CISS-driven reduction chemistry, especially on the icy mantle grain surfaces where complex organic molecules are likely formed, resulting in the chiral asymmetry of chiral molecules. For the ISM and star-forming regions, we found that the CDRC could produce an enantiomer excess of above $20\%$ for the reasonable range of spin-orbit coupling potential.

\item Magnetic aligned grains act as chiral agents due to chiral-induced spin-selective adsorption (CISA) effect and exchange (spin-spin) interaction, which amplifies the chiral asymmetry for complex molecules in the grain mantle surface. Since aligned grains are observed in a wide range of environments, from the diffuse ISM to star- and planet-forming regions, we expect a broad implication of SPEs and aligned grains for astrochemistry and astrobiology.


\item Our proposed mechanism for SPEs based on magnetically aligned grains is more universal than ferromagnetic deposits in the prebiotic Earth required for the proposal in \cite{Ozturk.2022} because grains are aligned with the magnetic field in a wide range of astrophysical environments and can work for any magnetic material (para-, superpara, and ferro-/ferri-magnetic material). If SPEs could induce an initial chiral asymmetry, our result implies that life would be more ubiquitous in the universe than previously thought

\item Our numerical estimates show that the disalignment of electron spins in elastic scattering by atoms/molecules due to the spin-orbit interaction is negligible due to much a longer Larmor timescale of the electron spin around the induced magnetic field than the collision timescale. Low-energy SPEs can be captured by protons in the ionized gas, but in the low ionization gas, SPEs may be captured by dust grains before protons. 
\item 
We also quantified the depolarization by SPEs by the dominant elastic scatteirng using the Mott scattering theory and found that the depolarization is rather small for low-energy SPEs.

\end{enumerate}

It is noted that \cite{Rosenberg.2019} discussed various examples where UV irradiation of magnetic surface in the ISM can result in SPEs. However, metallic surfaces may have random orientation in space due to gas collisions. As a result, direct irradiation of such randomly oriented magnetic surfaces would not lead to SPEs. Our proposed mechanism is based on aligned grains with the magnetic field, which can create SPEs. Since aligned grains are ubiquitous in a wide range of environments, including planet-forming disks where planets, comets, and asteroids are forming, resulting SPEs would enrich the chiral asymmetry of chiral molecules, leading to the homochirality of amino acids observed on Earth and meteorites.

\acknowledgments
We thank the anonymous referee for insightful comments that improve the paper. We thank Dr. Nguyen Xuan Dung (Institute of Basic Science, South Korea) for discussion on spin polarization of electrons by elastic scattering. This work is supported by the research planning for exploring cosmic life phenomena (LiCE) project (No. 2024E84100) funded by Korea Astronomy and Space Science (KASI). This work was also partly supported by a grant from the Simons Foundation to IFIRSE, ICISE (916424, N.H.). We would like to thank the ICISE staff for their enthusiastic support.

\appendix

\section{The CISS effect}\label{apdx:CISS}
This CISS effect was first discovered by \cite{Ray.1999} and \cite{Naaman.2012} where the authors found that the transmission of electrons through chiral molecules is spin-dependent. The later experiment by \cite{Gohler.2011} showed that transmitted electrons through doubled-stranded (chiral) DNA are spin-polarized, and photoelectrons are spin-polarized even when they are produced by unpolarized light. 

Although a quantitative theory of the CISS effect is not yet available (see \citealt{Evers.2022} for a review), the basic physics can be summarized as follows. Chiral molecules exhibit a helical electric field $\bE_{\rm helix}$ due to helical charge distribution (i.e., asymmetric charge distribution) caused by the molecule configuration \citep{Naaman.2012}. As the electron moves through a chiral molecule with velocity $\bv$, the electron experiences a centrepital, electrostatic force
\bea
\bF_{\rm cen}=-e\bE_{\rm helix},\label{eq:Fcen}
\ena
that causes electrons to change their direction and move along a helical trajectory. The helical motion of the electron with respect to the molecule axis induces an effective magnetic field, which is analogous to the electron orbital motion around the nucleus in an atom. In the electron's co-moving frame, electrons see the motion of the charge center of the chiral molecule, so that the electron is subject to an effective magnetic field caused by the charge center motion. The same consideration can be done using the Lorentz transformation between inertial frames. The interaction between the electron spin (magnetic moment) with the effective B-field induces the spin-orbit coupling (SOC, \citealt{Naaman.2012,Naaman.2018}), similar to spin-orbit coupling in atoms (see \citealt{RybickiLightman.1979}, p.252-253 for details).\footnote{In atoms, the spin-orbit coupling is due to the interaction of the electron spin (magnetic moment) and the magnetic field induced by the motion of atomic nucleus with respect to the electron.}

Following the special relativity of classical electromagnetism, the Lorentz transformation of magnetic fields from the molecule's frame to the electron co-moving frame is given by (see e.g., \citealt{RybickiLightman.1979}):
\bea
\bB_{\rm eff}=-\frac{\bv}{c}\times \bE_{\rm helix},
\ena
where $E,c,B$ are given in cgs units. To write in the SI units, one substitute $\bE\rightarrow \bE/c$ and the other parameters unchanged, which then becomes
\bea
\bB_{\rm eff}=-\frac{\bv}{c^{2}}\times \bE_{\rm helix},\label{eq:Beff}
\ena 
which implies that when the charged particle moves in an external electric field, there exists a magnetic field in the frame fixed to the charged particle.

The interaction between the electron's magnetic moment and the effective magnetic field induces an additional energy for the electron
\bea
V_{\rm SO}=-\bmu.\bB_{\rm eff}=\frac{g_{e}\mu_{B}}{\hbar}\bS.\bB_{\rm eff}.\label{eq:HSO_eff}
\ena
where $\bmu=\gamma_{e}\bS=\gamma_{e}\hbar \bsigma$ with $\gamma_{e}=-g_{e}\mu_{B}/\hbar$ is the gyromagnetic ratio with and $g_{e}\approx 2$ for electrons.\footnote{\cite{Naaman.2012} derived $\bB_{\rm eff}$ using the Hamiltonian SO coupling (after Eq. 3), and obtained $\bB_{\rm eff}=\bv/c^{2}\times \bE_{\rm helix}$, which leads to the potential $V_{\rm SO}=-g_{e}\mu_{B}(\bS/\hbar).{\bf B}_{\rm eff}$ instead of the correct one in Equation (\ref{eq:HSO_eff}). The correct sign is used in \cite{Ozturk.2022}.}

Therefore, the electron energy level is split the energy into three levels, depending on the spin direction $\bS$ (i.e., Zeeman effect).

Using $B_{\rm eff}$ from Equation (\ref{eq:Beff}) one obtains
\bea
V_{\rm SO}= -\frac{g_{e}\mu_{B}}{2\hbar c} \bS.(\bv\times \bE_{\rm helix})=-\frac{g_{e}\mu_{B}}{2\hbar c} \bE_{\rm helix}.[\bS\times \bv]=-\frac{g_{e}\mu_{B}}{2\hbar m_{e}c} \bE_{\rm helix}.[\bS\times \bp],\label{eq:HSO}
\ena
which corresponds to the Zeeman splitting of the electron energy into three different levels. Therefore, when passing through a chiral molecule, one electron spin state is favored while the other is disfavored, resulting in the electron spin-filtering effect. We note that our above formula is the same as the SO potential from the Dirac equation in \cite{RybickiLightman.1979} (see p.252-253), also as in \cite{Manchon.2015}. Yet, other authors (e.g., \citealt{Naaman.2012,Ozturk.2022}) assumed $V_{\rm SO}=\frac{g_{e}\mu_{B}}{2\hbar m_{e}c} \bS.[\bp\times \bE_{\rm helix}]$.

The magnitude of $V_{\rm SO}$ depends on the relative orientations of electron spin, electron momentum, and the molecule helical axis. However, the effect is always effective. Indeed, if initially the electron spin and momentum are aligned, then, the interaction potential is zero due to $\bS\times \bp=0$. As soon as the electron moves along the chiral field, its momentum is changed by the centripetal force and makes some angle with $\bS$, resulting in a non-zero $V_{\rm SO}$.

\section{Effect of grain magnetic moments on chiral molecules}\label{apdx:grain_chiral}

Consider silicate grains containing embedded iron clusters with the number of iron atoms per cluster, $N_{\rm cl}$, and the volume filling factor, $\phi_{\rm sp}$. The existence of iron clusters makes composite grains become superparamagnetic material that has magnetic susceptibility increased by a factor of $N_{\rm cl}$ from ordinary paramagnetic material \citep{DavisGreenstein.1951,JonesSpitzer.1967}. Therefore, the magnitude of Barnett magnetic moment becomes
\bea
\mu_{\rm Bar}\simeq 4.6\times 10^{-17}T_{g,1}^{1/2}a_{-5}^{1/2} St\left(\frac{N_{\rm cl,4}\phi_{\rm sp,-2}}{T_{d,1}}\right)~{\rm esu},~~~\label{eq:muBar_mag}
\ena
where $St$ is the suprathermal rotation parameter defined by $St=\Omega/\Omega_{T}$ and $\Omega_{T}=(kT_{\rm gas}/I_{a})^{1/2}$ with $I_{a}$ being the grain inertia moment, $a_{-5}=a/10^{-5}\cm$, $N_{cl,4}=N_{cl}/10^{4}$, $\phi_{\rm sp,-2}=\phi_{\rm sp}/10^{-2}$ where the normalization factor of $10^{-2}$ corresponds to $3\%$ of iron abundance embedded in the dust in the form of iron clusters (see \citealt{HoangLaz.2016,HoangBao.2023}), $T_{g,1}=T_{\rm gas}/10\K$ and $T_{d,1}=T_{d}/10\K$ with $T_{\rm gas}, T_{d}$ being the gas and dust temperatures.

The grain's magnetic moment produces a magnetic potential at a large distance $r$ from the dipole, which is given by (\citealt{Landau:1984ui})
\bea
\phi(\br)=\frac{\bmu.{\br}}{r^{3}},
\ena
and the magnetic field
\bea
{\bB}(\br)=-\nabla \phi(\br).
\ena

The energy potential due to dipole-dipole interaction of the magnetic grain and the chiral molecule of dipole moments $\bmu$ and $\bmu'$, separated by a distance $r$, is given by (see, e.g., \citealt{Nuth.1995,HoangBao.2023})
\bea
U_{dd}=-\bmu.{\bB}(\br)=(\bmu.\nabla)\phi(\br)
=\frac{\bmu.\bmu'}{r^{3}}-3\frac{(\bmu.{\br})(\bmu'.{\br})}{r^{5}}.\label{eq:Um_general}
\ena

For parallel or anti-parallel magnetic moments, the potential energy becomes 
\bea
U_{\rm dd}
=\frac{-\bmu.\bmu'}{r^{3}}(3\cos^2{\theta}-1),\label{eq:Um_para}
\ena
where $\theta$ is the relative angle between $\br$ and $\bmu$. 

For parallel magnetic dipoles, the grain Barnett dipole interaction is attractive at its maximum of $U_{\max} = -2\mu.\mu'/r^{3}$ for $\theta=0^{\circ}$ and repulsive for $\cos\theta<1/\sqrt{3}$. For anti-parallel dipoles, the interaction is attractive at its maximum for $\theta=90^{\circ}$ and repulsive for $\theta=0$. Therefore, the dipole-dipole interaction favors the attraction along the alignment axis for parallel spins and the attraction along the perpendicular direction for anti-parallel spins.

Equation (\ref{eq:Um_general}) implies the interaction potential between the aligned grain of magnetic moment $\mu_{\rm Bar}$ (Eq. \ref{eq:muBar_mag}) and a chiral molecule of magnetic moment $\bmu_{\rm mol}$ as given by
\bea
U_{dd}=\frac{-\bmu_{\rm Bar}.\bmu_{\rm mol}}{r^{3}}(3\cos^{2}\theta-1)
\simeq-0.027St_{2}r_{0.01\mum}^{-3}p_{\rm mol,3}(3\cos^{2}\theta-1) \hat{\bS}_{d}.\hat{\bS}_{\rm mol}~ \ev,~~~\label{eq:Uchiral}
\ena
where $St_{2}=St/10^{2}$, $p_{\rm mol}=\mu_{\rm mol}/\mu_{B}$ is the strength of molecule magnetic moment and $p_{mol,3}=p_{\rm mol}/10^{3}$, $\hat{\bS}_{d}$ is the unit vector of the grain spin (angular velocity) and $\hat{\bS}_{\rm mol}$ is the unit vector of molecule spin.

Equation (\ref{eq:Uchiral}) shows that the dipole-dipole interaction potential can be larger than the kinetic energy of chiral molecules of $kT_{\rm gas}=0.0013(T_{\rm gas}/15\K)$ eV for typical cold clouds of $T_{\rm gas}=15\K$. Thus, they tend to attract chiral molecules with similar spins and repel the other due to magnetic dipole-dipole (or spin-spin) interaction, facilitating the formation of more complex molecules of similar chirality and increasing the chiral asymmetry.



\bibliography{ms.bbl}
\end{document}